\documentclass[12pt]{article}
\usepackage{epsfig}
\usepackage{amsfonts}
\usepackage{amssymb}
\usepackage{amsmath}
\usepackage{ifthen}
\newcommand\cfdplus{CFD$^+$}

\newcommand\eg{e.g.}
\newcommand\ie{i.e.}
\newcommand\etal{\textit{et al}}
\newcommand\itoc{\ensuremath{\mathsf{I2C}}}

\newcommand*{\qed}{\hfill\ensuremath{\blacksquare}}

\newtheorem{definition}{Definition}
\newtheorem{proposition}{Proposition}
\newtheorem{theorem}{Theorem}
\newtheorem{lemma}{Lemma} 
\newtheorem{remark}{Remark} 
\newtheorem{corollary}{Corollary}

\newenvironment{proof}{\paragraph{Proof:}}{\hfill$\square$}

\newenvironment{defin}[1][]{\ifthenelse{\equal{{\bf #1}}{}}{\definition}{\definition[#1]}\rm}{\enddefinition}

\newenvironment{propo}[1][]{\ifthenelse{\equal{#1}{}}{\proposition}{\proposition[#1]}\rm}{\endproposition}
\newenvironment{remar}[1][]{\ifthenelse{\equal{#1}{}}{\remark}{\remark[#1]}\rm}{\endremark}
\newenvironment{theo}[1][]{\ifthenelse{\equal{#1}{}}{\theorem}{\theorem[#1]}\rm}{\endtheorem}
\newenvironment{corol}[1][]{\ifthenelse{\equal{#1}{}}{\corollary}{\corollary[#1]}\rm}{\endcorollary}
\newenvironment{lem}[1][]{\ifthenelse{\equal{#1}{}}{\lemma}{\lemma[#1]}\rm}{\endlemma}

\newcommand\figref[1]{Figure \ref{#1}}
\newcommand\defref[1]{Definition \ref{#1}}
\newcommand\propref[1]{Proposition \ref{#1}}
\newcommand\theoref[1]{Theorem \ref{#1}}
\newcommand\lemref[1]{Lemma \ref{#1}}
\newcommand\secref[1]{Section \ref{#1}}

\newcommand\xxx[1]{\ensuremath{\mathsf{#1}}}

\newcommand\veh{\xxx{vehicle}}
\newcommand\ve{\xxx{veh}}

\newcommand\eng{\xxx{engine}}
\newcommand\en{\xxx{en}}

\newcommand\gear{\xxx{gear}}
\newcommand\ger{\xxx{ge}}

\newcommand\auto{\xxx{automatic}}
\newcommand\aut{\xxx{a}}

\newcommand\manu{\xxx{manual}}
\newcommand\man{\xxx{m}}

\newcommand\brake{\xxx{brake}}
\newcommand\br{\xxx{br}}

\newcommand\gas{\xxx{gas}}
\newcommand\g{\xxx{g}}

\newcommand\elec{\xxx{electric}}
\newcommand\el{\xxx{el}}

\newcommand\wheel{\xxx{wheel}}
\newcommand\w{\xxx{w}}

\newcommand\axle{\xxx{axle}}
\newcommand\ax{\xxx{ax}}

\newcommand\abs{\xxx{abs}}
\newcommand\ab{\xxx{ab}}

\newcommand\xxxf{\xxx{f}}
\newcommand\xf[1]{\xxx{f_{#1}}}

\newcommand\prnt[1]{\ensuremath{#1^\uparrow}}
\newcommand\chld[1]{\ensuremath{#1_\downarrow}}

\newcommand\gchld[1]{\ensuremath{#1_{\downarrow\downarrow}}}
\newcommand\bfanfm{\bfstrname{M}}

\newcommand\inducN[2]{{#1}^{#2}}
\newcommand\inducD[2]{{#1}_{-{#2}}}

\newcommand\products{\setofsetname{P}}
\newcommand\pproducts{\setofsetname{P}^\mathrm{flat}}

\newcommand\cfdfam{\setofsetname{D}}

\newcommand\agrp{\setofsetname{G}}

\newcommand\ancd{\bfstrname{D}}
\newcommand\acrd{\setofsetname{C}}
\newcommand\asol{\setofsetname{S}}

\newcommand\depth[1]{depth({#1})}

\newcommand\biimplies{\Longleftrightarrow}

\newcommand\lra{\longrightarrow}

\newcommand\rarr{\rightarrow}

\newcommand\nat{\mathbb{N}}

\newcommand\eqdef{\ensuremath{\stackrel{\mathrm{def}}{=}}}

\newcommand\restr[2]{{ \left.\kern-\nulldelimiterspace #1 \vphantom{\big|} \right|_{#2} 
  }}

\newcommand\bfstrname[1]{\ensuremath{\mathbf{#1}}}

\newcommand\setofsetname[1]{\ensuremath{\mathcal{#1}}}

\newcommand\multisets[1]{\setofsetname{MS}(#1)}
\newcommand\multhier{\setofsetname{H}}
\newcommand\treemulthier{\setofsetname{TH}}

\newcommand\pure[2]{flat_{#1}({#2})}
\newcommand\relaxm[1]{{#1}^\circ}

\newcommand\prntmult[2]{\ensuremath{#2^{\uparrow_{#1}}}}
\newcommand\multing[1]{MsIng(#1)}

\newcommand\rankm[1]{ rank({#1}) }
\newcommand\rootm[1]{ root({#1})}
\newcommand\dom[1]{ dom({#1})}

\newcommand\treem[1]{ T_{#1} }
\newcommand\mergcfd[1]{ \ancd_{\merg{#1}} }
\newcommand\mergcfds[1]{ \setofsetname{D}_{\merg{#1}} }

\newcommand\treefam{\setofsetname{T}}
\newcommand\merg[1]{{#1}^{\mathrm{merge}}}

\def\beq{\begin{equation}}
\def\eeq#1{\label{#1}\end{equation}}
\def\eeqn{\end{equation}}

\def\beqa{\begin{eqnarray}}
\def\eeqa#1{\label{#1}\end{eqnarray}}
\def\eeqan{\end{eqnarray}}

\let\bar=\overbar

\def\etal{{\it et al.}}
\def\ie{{\it i.e.}}
\def\eg{{\it e.g.}}

\def\Dslash{\not{\hbox{\kern-4pt $D$}}}
\def\dslash{\not{\hbox{\kern-2pt $\del$}}}

\def\msb{{\bar{\ssstyle M \kern -1pt S}}}

\def\Title#1{\begin{center} {\Large {\bf #1} } \end{center}}

\begin{document}

\Title{Multiset Theories of Cardinality-based Feature Diagrams}

\bigskip\bigskip


\begin{center}  
 Aliakbar Safilian\index{Safilian, A.} and Tom Maibaum\index{Safilian, A.}\\
{\it Department of Computing and Software, 
McMaster University, Canada}
\bigskip\bigskip
\end{center}

\begin{abstract}  
Software product line engineering is a very common method for designing complex software systems. Feature modeling is the most common approach to specify product lines. 
The main part of a feature model is a special tree of features called a feature diagram. Cardinality-based feature diagrams provide the most expressive tool among the current feature diagram languages. The most common characterization   of the semantics of a cardinality-based diagram is the set of flat multisets over features satisfying the constraints. However, this semantics provides a poor abstract view of the diagram. We address this problem by proposing another multiset theory for the cardinality-based feature diagram, called the hierarchical theory of the diagram. We show that this semantics  captures all information of the diagram so that one can retrieve the  diagram from its hierarchical semantics.    We also characterize sets of multisets, which can provide a  hierarchical semantics of some diagrams. 
\end{abstract}

\section{Introduction}
{\em Product line} (PL) engineering \cite{pohl2005} is a popular method of designing complex software/hardware systems. There are  many successful industrial stories applying product line engineering, \eg, ``Mega-Scale Product Line Engineering at General Motors'' \cite{flores2012}, ``HomeAway case study'' \cite{krueger2008}, ``LG Industrial Systems'' \cite{pohl2005}, ``Lufthansa Systems'' \cite{chastek2011}, and ``Nokia Mobile Phones, Browsers, and Networks'' \cite{maccari2002, Jazayeri2000, jaaksi2002}.  A PL is a set of products that share some commonalities along with having some variabilities, where commonalities and variabilities are usually captured using entities called {\em features}, system properties that are relevant to some stakeholders  \cite{czarnecki2005}. The idea of PL engineering is that, instead of producing products individually, the common core of a product line is produced, leaving a much smaller task to be completed, namely the adaptation of the core to a concrete application requirement. The advantages of the method include significant reduction in development time and cost \cite{pohl2005}, reusability \cite{bosch2001}, reduced product risks \cite{quilty2011}, and increased product quality \cite{jensen2007}. 

 {\em Feature modeling} is a common approach for modeling PLs. The main part of a feature model, called a {\em feature diagram} (FD), is  a tree of features equipped with some special annotations on the tree's elements showing the {\em constraints} on features based on which valid configurations are built.  FDs are grouped into {\em basic} FDs (BFDs) and {\em cardinality-based} FDs (CFDs).   
\begin{figure}[h]
\centering
    \includegraphics[scale=0.45]{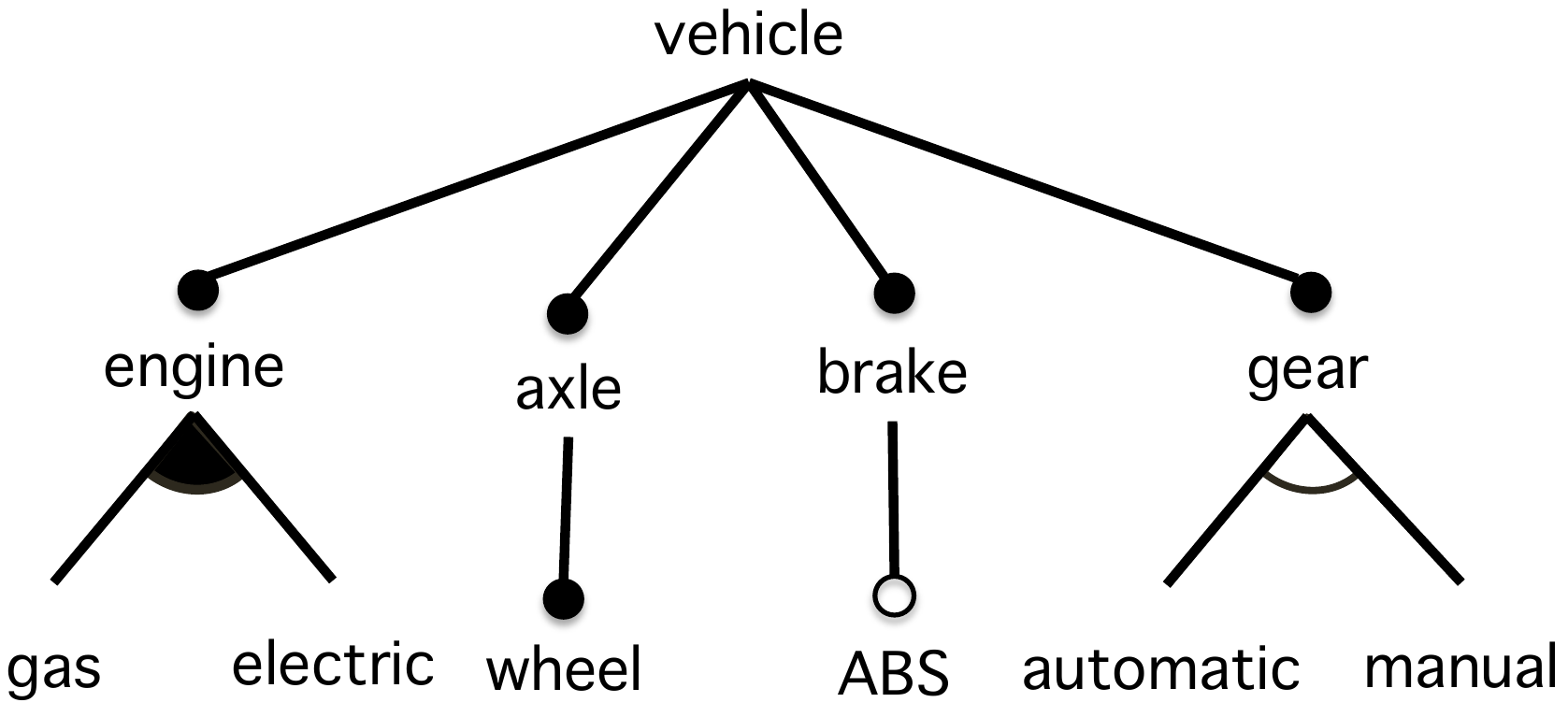}
\caption{A Basic Feature Diagram}
\label{fig:BFD-intro}
\end{figure}

BFDs represent product variability and commonality in terms of Boolean constraints: {\em optional/mandatory} features, and {\em OR/XOR} {\em groups}.      \figref{fig:BFD-intro} is a BFD of a part of a vehicle system:  An edge with a black circle shows a  mandatory feature: every  \veh\ must include \eng, \axle, \brake, and  \gear; an axle must include \wheel. An edge with a hollow circle shows an  optional feature:  a \brake\ can be optionally equipped with \abs. These two types of edges (mandatory and optional) are called {\em solitary}.  Black angles denote OR groups:  the OR group \{\gas, \elec\} indicates that an engine can be either gasoline or electric, or both. XOR groups are usually shown by hollow angles: the XOR group \{\auto, \manu\} states that a gear can be either automatic or manual, but not both. 
\begin{figure}[h]
\centering
    \includegraphics[scale=0.5]{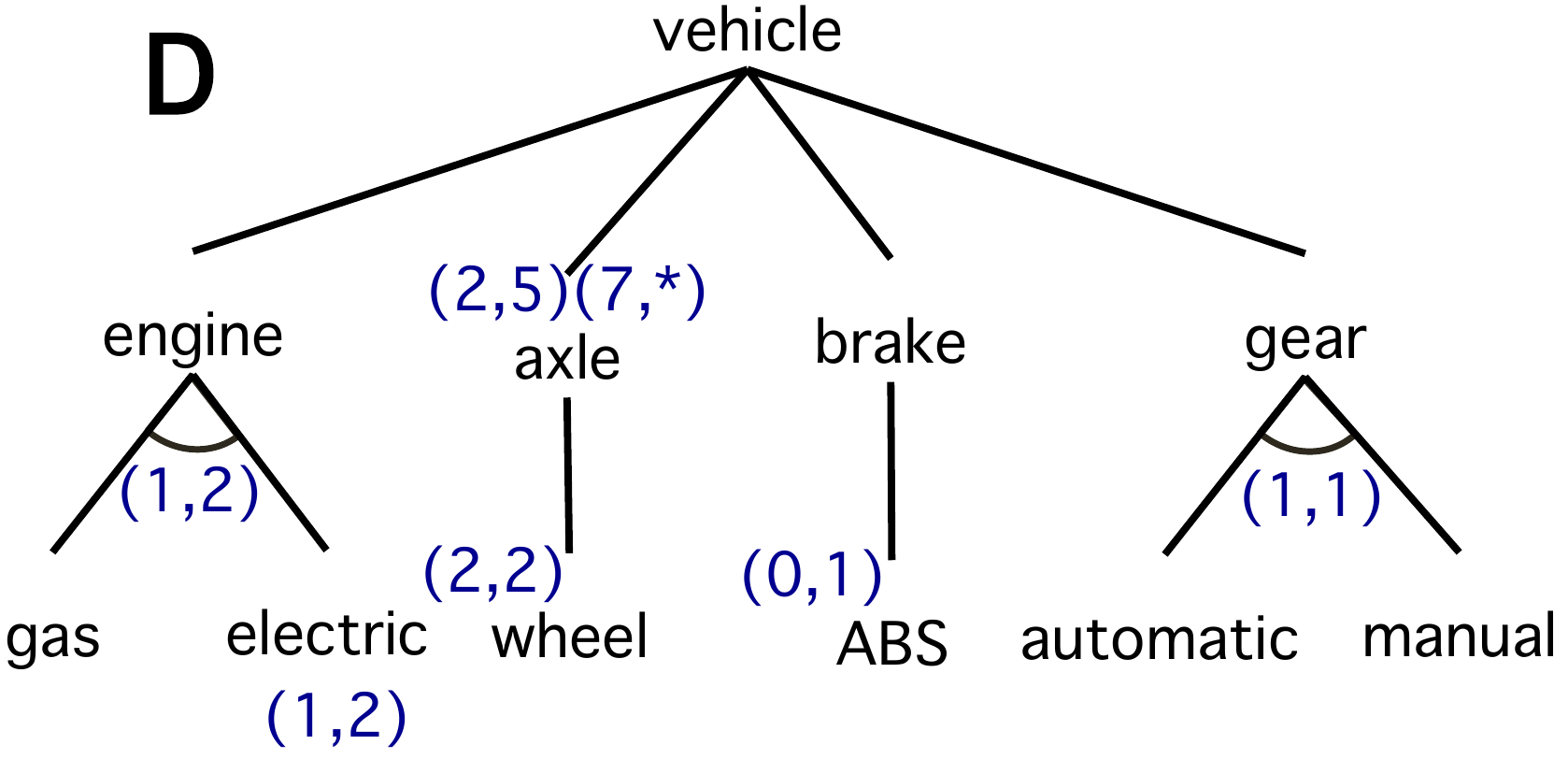}
\caption{A cardinality-based feature diagram}
\label{fig:CFD-intro}
\end{figure}

In CFDs, multiplicities are used in place of traditional Boolean annotations. There are two types of multiplicity constraints: {\em feature} and {\em group} multiplicity constraints. A multiplicity constraint is usually expressed as a sequence of pairs $(l,u)$, where $l$ is a  natural number, $u$ is either a number or $*$ (representing an unbounded multiplicity) and $l\leq u$. Henceforth, we call a multiplicity constraint on a node or group a {\em multiplicity domain}. \figref{fig:CFD-intro} provides a CFD for a vehicle system. As a common convention, we assume the multiplicity domain $(1,1)$ on an element (either a feature or a group)  if no multiplicity domain is shown on the element. A vehicle has 4 subfeatures: \eng\ with multiplicity $(1,1)$ (a vehicle must have engine), \axle\ with multiplicity domain $(2,5)(7,*)$ (a vehicle can have any number of axles excluding 0, 1, and 6), \brake\ with multiplicity domain $(1,1)$ (a vehicle must have exactly one brake), and \gear\ with multiplicity domain $(1,1)$ (a vehicle must have exactly one gear).   The children of \eng\ are grouped with multiplicity domain $(1,2)$ meaning that an engine can be either electric or gasoline, or both. The multiplicity domain $(1,1)$ on \gas\ indicates that a vehicle can have at most 1 gasoline engine if it has any gasoline engine. The multiplicity domain $(1,2)$ on \elec\ means that a vehicle can have at most 2 electric engines if it has any.  The multiplicity constraint $(2,2)$ on \wheel\ denotes that an axle must have exactly 2 wheels. The feature multiplicity domain $(0,1)$ on \abs\ models its optional presence in a brake.  The subfeatures of \gear\ are grouped with multiplicity domain $(1,1)$, which means that a gear can be either automatic or manual, but not both. The multiplicity domains of \auto\ and \manu\ are both $(1,1)$. 

We may also want to add some constraints involving incomparable features. (Two features of a given FD are called incomparable if neither of them is a descendant of the other in the feature diagram.) Such constraints are called {\em crosscutting constraints} (CC) (a.k.a. additional constraints). For an example, let ``\manu\  includes \abs'' be a CC over our examples. It states that a vehicle with a manual gear must have an ABS brake.

As we noticed, the difference between basic  and cardinality-based feature modeling is that the former deals with feature ``types'', while we deal with feature ``resources'' (occurrences) in the latter. CFDs are more expressive than BFDs \cite{czarnecki2005}, as any Boolean constraint can be expressed in terms of multiplicities. Henceforth, {\bf we deal with CFDs}.   

The most common semantics considered for a CFD in the literature is the set of its valid  {\em flat configurations}\footnote{Usually called {\em products} in the literature\cite{czarnecki2005}.}, where a flat configuration of a CFD is a multiset of features satisfying the constraints of the CFD  \cite{czarnecki2005}.  For  example, the multiset $\lceil \veh, \eng, \gas,  \axle^2, \wheel^4, \brake,  \gear, \manu  \rceil$\footnote{We use the symbols $\lceil, \rceil$ as multiset identifiers-- see page \pageref{pp:multiset} for a formal definition of mutlisets.} is a valid flat configurations of the CFD in  \figref{fig:CFD-intro}. We call this semantics the {\em flat semantics} of CFDs.  
\begin{figure}[h]
\centering
    \includegraphics[scale=0.4]{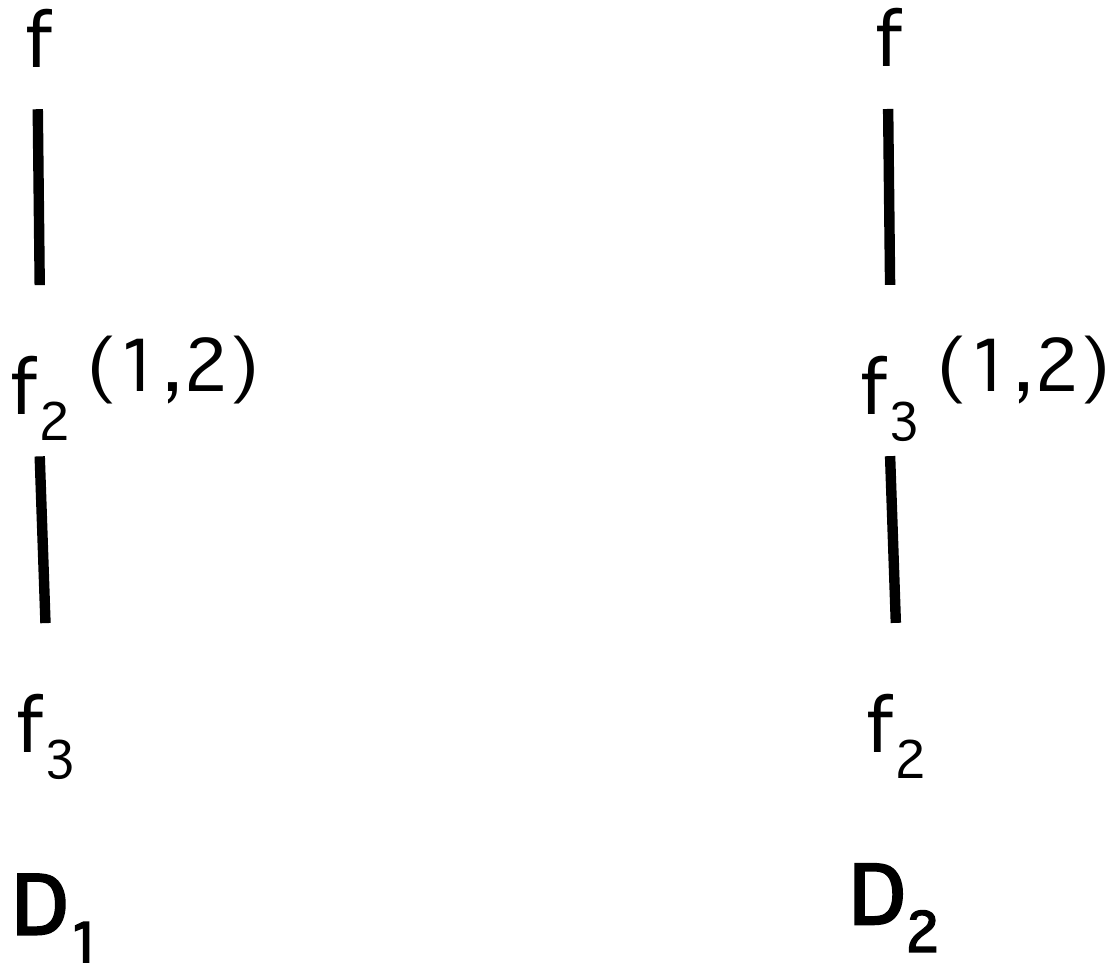}
\caption{Two different CFDs with the same flat semantics}
\label{fig:2-flat-equi}
\end{figure}

The flat semantics of a CFD provides a useful abstract view of the CFD, as it can address a large number of analysis questions about the CFD, including ``decide whether a given multiset is a valid product of a given CFD or not'', ``decide whether a given integer is a valid multiplicity of a given feature or not'', etc. However, it is a poor abstract view, as it does not capture some other useful information about the CFD, such as the {\em hierarchical structure}.   For an example, consider two different CFDs $\ancd_1$ and $\ancd_2$ in \figref{fig:2-flat-equi}. They are equivalent in the flat semantics, since they represent the same flat products $ \lceil \xxxf, \xf{2}, \xf{3} \rceil$ and $\lceil \xxxf, \xf{2}^2, \xf{3}^2 \rceil$. Therefore, the analysis questions relying on the lost information, including the {\em least common ancestor} of a given set of features, {\em root} feature, {\em subfeatures} of a given feature \cite{benavides2010},  cannot be addressed using the flat semantics. There are some other analysis issues, in which the use of a poor abstract view can be error-prone. For example, it is often important to know if one feature model  is a {\em refactoring} of another feature model, or one is a {\em specialization} of another, or neither \cite{thum2009}. These questions are semantics based. Relying on a poor semantics to define such analysis questions makes the definitions poor for their goals. Another deficiency of a poor semantics is relevant to {\em reverse engineering} of feature models. Indeed, the main reason making the current state of the art approaches \cite{she2011, lopez2015} heuristic is caused by using such a poor abstract view of FDs. 

In this paper, we address this problem by proposing another multiset based semantics of CFDs.  To this end, we first define a {\em hierarchy of multisets} built over features.  Then, we define a {\em hierarchical product} of a CFD as a multiset (in the corresponding multisets hierarchy) such that its rank in the hierarchy corresponds to the depth of the CFD. The set of all hierarchical products is called the {\em hierarcical multiset theory} (or semantics) of the CFD. We then prove that the hierarchical theory of a given CFD captures {\em all} information about the diagram so that one can get back to the diagram from its hierarchical theory. 
 
 The structure of the rest of the paper is as follows:  \secref{sec:cfd-flat} starts with our formal framework for the syntax of CFDs. In this section, we also provide a recursive definition for flat semantics of CFDs.  \secref{sec:hier} discusses and formalizes the idea of hierarchical products for CFDs.  \secref{sec:tree-mult} characterizes multisets representing hierarchical products of some CFDs. To this end, we introduce the notion of {\em tree-like multisets}. It is proven that a multiset is a hierarchical product of a given CFD iff it is a tree-like multiset. \secref{sec:merge-tree-mult} characterizes  sets of tree-like multisets representing hierarchical semantics of  CFDs, namely, we show what sets of tree-like multisets are the hierarchical semantics of some CFDs.  To this end, the notion of {\em mergeable} and {\em complete mergeable} tree-like multisets are introduced. We discuss some other practical applications of the work in \secref{sec:practice}.   Related work is discussed in \secref{sec:related}. \secref{sec:conclusion} concludes the paper with a discussion, other practical applications of the hierarchical multiset semantics, and future work.   Proofs of the main theorems along with some other complementary definitions and lemmas can be found in the Appendix.

\section{CFDs and Flat MSet Theories} \label{sec:cfd-flat}

 A CFD is a tree of features in which some subsets of non-root nodes are {\em grouped} and other nodes are called {\em solitary}. In addition, non-root nodes and groups are equipped with some multiplicity constraints.  As discussed in the Introduction, a multiplicity  domain is usually expressed as a {\em finite} sequence  $(l_1,u_1) \ldots (l_n, u_n)$, where $l_i$ ($\forall 1\leq i \leq n$) is a  natural number, $u_i$ ($\forall 1\leq i \leq n$) is either a number or $*$ (an unbounded multiplicity),  $l_i \leq u_i$ (we assume $k\leq *$ for any number $k$), and $u_i \leq l_{i+1}$ ($\forall 1 \leq i < n$)\footnote{See \cite{safilian2015, czarnecki2005} for some detailed definitions.}. However, a multiplicity domain can be expressed as a subset of natural numbers, \eg, the multiplicity domains (2,5)(7,*) and (1,2) on the feature $\axle$ and the group $\xxx{G} = \{\gas, \elec\}$ in \figref{fig:CFD-intro} are the sets $\nat \setminus \{0, 1, 6\}$ and $\{1,2\}$, respectively. In this paper, we consider this form of definition, \ie, subsets of natural numbers, for  multiplicity domains. This definition of multiplicity domains makes some further formalizations in the paper easier to read. 
 \begin{remar}
 Note that considering any subset of natural numbers as a valid multiplicity domain makes CFDs more expressive than traditional CFDs, as not all subsets of natural numbers can be expressed as a finite sequence of intervals, \eg, the set of even numbers. 
 \qed
 \end{remar}

 The following definition formalizes the syntax of CFDs. In our framework, solitary nodes are derived constructs. In the following definition $\nat$ denotes the set of natural numbers. 
\begin{defin}[Cardinality-based Feature Diagrams]\label{def:CFD-mult}
 A {\em cardinality-based feature diagram} (CFD) is a 5-tuple $\ancd=(F, r, \_^\uparrow, \agrp, \acrd)$, where:

(i) $T=(F, r, \_^\uparrow)$  is a {\em tree} with set $F$ of nodes (called {\em features}), $r\in F$ is the root, and  function $\_^\uparrow$ maps each non-root node $f\in F_{-r}\eqdef F\setminus \{r\}$ to its parent \prnt{f}.  The inverse function that assigns to each node $f$ the set of its children is denoted by $\chld{f}$. The set of all descendants of  $f$ is denoted by  $\gchld{f}$.

(ii) $\agrp \subseteq 2^{F_{-r}}$ is a set of {\em grouped} nodes.  For all $G\in\agrp$, $|G| > 1$,   and all nodes in $G$ have the same parent, denoted by $\prnt{G}$. All groups in $\agrp$ are disjoint, \ie, $\forall G, G' \in \agrp: (G = G') \vee (G \cap G' = \varnothing)$.  The nodes that are not in a group are called  {\em solitary} nodes. Let \asol\ denote the solitary nodes, \ie, $\asol = F_{-r} \setminus \bigcup_{G\in \agrp} G$.

(iii) $\acrd: (F_{-r} \cup \agrp)  \rarr 2^\nat$ is a total function called the  {\em multiplicity function}. For any feature or group $e\in F_{-r} \cup \agrp$, $\acrd(e)$ represents the multiplicity constraint of $e$, where $\acrd(e) \neq \{0\}$ and $\acrd(e) \neq \varnothing$. In addition, for all $G\in \agrp$, $\acrd(G)$ is a finite subset of $\nat$ and its greatest member is less than or equal to $|G|$ (the cardinality of $G$). 

The class of all  CFDs over the same set of features $F$ is denoted by $\cfdfam(F)$. If needed, we will subscript \ancd's components with index $_\ancd$, \eg, write $\agrp_\ancd$.  \qed
 \end{defin}

\label{pp:multiset} A multiset over a set $F$ is a total function $m: F \rarr \nat$.  For any $f\in F$, $m(f)$ is called the multiplicity of $f$ in $m$. The set $\{f\in F: m(f) > 0\}$ is called the {\em domain} of $m$, denoted by $\dom{m}$. The multiset $m$ is called finite if $\dom{m}$ is finite.  We also need the {\em additive union} operation, denoted by $\uplus$, on multisets:  $(m\uplus m')(f)  = m(f) + m'(f)$. We write $m = \lceil a_1^{n_1}, a_2^{n_2}, \ldots \rceil$ to explicitly show the elements of a multiset $m$, where $n_i = m(a_i)$ for any  $a_i\in \dom{m}$. We usually do not write a multiplicity 1 on an element in a multiset, \eg, $\lceil a, b^2\rceil = \lceil a^1, b^2 \rceil$. The empty multiset is denoted by $\varnothing$.   

A {\em flat product} of a given CFD is a multiset of features satisfying the constraints of the CFD: (i) the root is in the domain of the multiset with multiplicity 1, \eg, $m(\veh) =1$ for any valid flat product $m$ of the CFD in \figref{fig:CFD-intro};  (ii) for a non-root features $f$ included in the flat product, there must be a multiplicity $c$ in $f$'s multiplicity domain such that its multiplicity in the flat product is equal to the product of its parent's  multiplicity and $c$, \eg, if the multiplicity of \axle\ in a flat product of the CFD in \figref{fig:CFD-intro} is 3 then \wheel's multiplicity must be 6; (iii) if the parent of a {\em mandatory feature} (a solitary feature with lower bound multiplicity greater than $0$) is included in a flat product then it must be included too, \eg, the presence of \axle\ in a flat product implies the presence of \wheel\ in the flat product; (iv) if a parent of a grouped set of features is included in a flat product then the presence of the grouped features must satisfy the associated group multiplicity constraint, \eg, the presence of \gear\ in a flat product implies the presence of either \manu\ or \auto\ in the flat product.\footnote{\cite{safilian2015} considers another condition as follows: If a non-root feature is included in a flat product then its parent must be included too. However, it is a consequence of our condition (ii).} 

For example, the following multisets are valid flat products of the CFD \ancd\ in \figref{fig:CFD-intro}, where $m = \lceil \veh, \gear, \brake, \eng \rceil$.
\label{pp:flatp}
$$m_1 = m \uplus \lceil \gas, \axle^3, \wheel^6, \manu  \rceil.$$
$$m_2 = m \uplus \lceil \elec, \axle^3, \wheel^6, \manu  \rceil.$$
$$m_3 = m \uplus \lceil \elec^2, \axle^3, \wheel^6, \manu  \rceil.$$ 
$$m_4 = m \uplus \lceil \gas, \elec, \axle^3, \wheel^6, \manu  \rceil.$$ 
$$m_5 = m \uplus \lceil \gas, \elec^2, \axle^3, \wheel^6, \manu  \rceil.$$
$$m_6 = m \uplus \lceil \gas, \axle^3, \wheel^6, \auto  \rceil.$$
$$m_7 = m \uplus \lceil \elec, \axle^3, \wheel^6, \auto \rceil.$$
$$m_8 = m \uplus \lceil \elec^2, \axle^3, \wheel^6, \auto  \rceil.$$ 
$$m_9 = m \uplus \lceil \gas, \elec, \axle^3, \wheel^6, \auto  \rceil.$$ 
$$m_{10} = m \uplus \lceil \gas, \elec^2, \axle^3, \wheel^6, \auto  \rceil.$$ 
$$m_{11} = m_1 \uplus \lceil \abs  \rceil.$$
$$m_{12} = m_2 \uplus \lceil \abs  \rceil.$$
$$m_{13} = m_3 \uplus \lceil \abs  \rceil.$$
$$m_{14} = m_4 \uplus \lceil \abs  \rceil.$$
$$m_{15} = m_5 \uplus \lceil \abs  \rceil.$$
$$m_{16} = m_6 \uplus \lceil \abs  \rceil.$$
$$m_{17} = m_7 \uplus \lceil \abs  \rceil.$$
$$m_{18} = m_8 \uplus \lceil \abs  \rceil.$$
$$m_{19} = m_9 \uplus \lceil \abs  \rceil.$$
$$m_{20} = m_{10} \uplus \lceil \abs  \rceil.$$

The following multisets are not valid flat products of the CFD: 

--$m \uplus \lceil \gas, \manu, \axle^3, \wheel^4  \rceil$ violates (ii) 

-- $m \uplus \lceil \gas, \manu, \axle^3 \rceil$ violates (iii) 

-- $m \uplus \lceil \gas, \axle^3, \wheel^6  \rceil$ violates (iv) \\

We formalize the flat products of a given CFD as follows: 
\begin{defin}[Flat Products] \label{def:plain-product}
Let $\ancd = (F, r, \_^\uparrow, \agrp, \acrd)$ be a CFD.  A  multiset $m$ over $F$ is called a {\em flat product} of $\ancd$ if  the following conditions hold:

(i) $m(r) = 1$,

(ii) $\forall f\in F_{-r}: f \in \dom{m} \implies (\exists c \in \acrd(f): m(f) = c \times m(\prnt{f}))$,

(iii) $\forall f\in \asol: 0\not\in\acrd(f)  \wedge m(\prnt{f}) > 0 \implies m(f) > 0$,

(iv) $\forall G\in \agrp: (m(\prnt{G}) > 0) \implies (|dom(m) \cap G| \in \acrd(G))$.
\\
The set of flat products of \ancd, denoted by $\pproducts(\ancd)$, is called the {\em flat theory} of \ancd. \qed
\end{defin}

We also provide a recursive definition of flat products in \lemref{th:flat-products}.  To this end, we first define a notion called {\em flat products associated with groups}. A flat product associated with a group is a multiset over the group's features satisfying the group's multiplicity domain. For example, consider the group $\xxx{G} = \{\gas, \elec\}$ in the CFD in \figref{fig:CFD-intro}. The set of hierarchical products associated with \xxx{G} consists of the following elements: $\lceil \gas \rceil$, $\lceil \elec \rceil$, $\lceil \elec^2 \rceil$, $\lceil \gas, \elec \rceil$, and $\lceil \gas, \elec^2 \rceil$. 
\begin{defin}[Flat Products Associated with Groups]\label{def:gp-products}
Let $\ancd = (F, r, \_^\uparrow, \agrp, \acrd)$ be a CFD and $G = \{f_1, f_2, \ldots, f_k\}\in \agrp$ for $k\in \nat$. A   multiset $m$ over $F$ is a {\em flat product} associated with $G$ if there exist $c\in \acrd(G)$, $c_i \in \acrd(f_i)$, $g_i \in \{0,1\}$, and $m_i\in \pproducts(\inducN{\ancd}{f_i})$ for any $1\leq i \leq k$ such that 
$$ 
m = \biguplus_{1\leq i \leq k}{m_i^{c_i\times g_i}}, \text{ and } \sum_{1\leq i \leq k}{g_i} = c
$$
The set of all flat products associated with a group $G$ is denoted by $\pproducts(\ancd, G)$. \qed
\end{defin}
 
 The following lemma provides a recursive definition for  the flat products of a given CFD.  
\begin{lem} \label{th:flat-products}
Given a CFD $\ancd = (F, r, \_^\uparrow, \agrp, \acrd)$, for any multiset $m$ over $F$:  $m \in \pproducts(\ancd)$ iff $m$ satisfies the following conditions:

(i)  $m(r) = 1$,

(ii) $\forall f\in \asol \cap \chld{r},~ \exists c\in \acrd(f),~ \exists n\in \pproducts(\inducN{\ancd}{f}),~ \forall e\in \dom{n}: m(e) = c\times n(e)$.

(iii) $\forall G\in \agrp \cap 2^{\chld{r}},~ \exists n \in \pproducts(\ancd, G),~ \forall e\in \dom{n}: m(e) = n(e)$.
\qed
\end{lem}
The following statement is a corollary of the above lemma. 
\begin{corol} \label{co:flat}
Given a CFD $\ancd = (F, r, \_^\uparrow, \agrp, \acrd)$, a flat product $m\in \pproducts(\ancd)$ satisfies the following conditions:
 
 (i) $\forall f\in \asol,~ \exists c\in \acrd(f),~ \exists n\in \pproducts(\inducN{\ancd}{f}),~ \forall e\in \dom{n}: m(e) = n(e) \times c \times m(\prnt{f})$.
 
(ii) $\forall G\in \agrp,~ \exists n \in \pproducts(\ancd, G),~ \forall e\in \dom{n}: m(e) = n(e) \times m(\prnt{G})$.
\qed
\end{corol}
As discussed in the Introduction, the flat semantics of a CFD provides a useful abstract view of the CFD. However, it is a poor abstract view of the CFD, as it does not capture some useful information about the diagram, such as the hierarchical structure. To address this problem, we propose another multiset theory for CFDs in the next section.


\section{Hierarchical MSet Theory of CFDs}\label{sec:hier}
Two types of information are lost in the flat theory of a CFD: the {\em tree structure}, and the {\em feature's types} (grouped or solitary).  For an example, given a CFD, we cannot address the following questions via the CFD's flat theory: what are the subfeatures of a given feature?  decide whether a given feature is solitary or not? In this section, we address this problem with another multiset theory for CFDs, called {\em hierarchical theory}. 

\begin{figure}[h]
\centering
    \includegraphics[scale=0.6]
                 {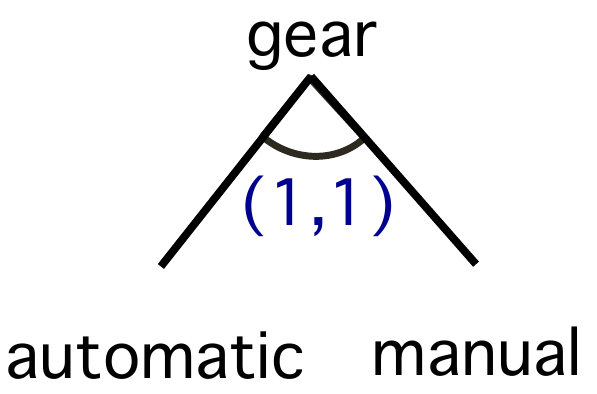}
\caption{Diagram induced by \gear\ in \ancd\ in \figref{fig:CFD-intro}}
\label{fig:induce-gear}
\end{figure}

We need the notion of {\em induced diagrams} to proceed. Given a CFD \ancd\ and a feature $f$, the diagram induced by $f$ is a CFD whose tree is the tree under $f$ in \ancd's tree and all other components are inherited from $\ancd$. For an example, the diagram induced by $\gear$  of the CFD $\ancd$ in \figref{fig:CFD-intro}, would be the CFD $(\{\gear, \manu, \auto\}, \gear, \prnt{\_}, \{\xxx{G} = \{\manu, \auto\}\}, \acrd)$, where $\prnt{\manu} = \prnt{\auto} = \gear$,  $\acrd(\xxx{G}) = \{1\}$, $\acrd(\manu) = \acrd(\auto) = \{1\}$. The corresponding CFD is represented in \figref{fig:induce-gear}. 

The notion induced diagrams by nodes is formalized in the following definition.  For a relation $R\subseteq B \times C$ and a set $A$,  the notation $\restr{R}{A}$  is used to denote the restriction of $R$ to $A$.

\begin{defin} [Diagrams Induced by Nodes]  \label{def:cfm-induce}
Let $\ancd = (F, r, \prnt{\_}, \agrp, \acrd)$ be a CFD and $f\in F$. The {\em CFD induced by $f$} is a CFD $\inducN{\ancd}{f} = (F', f, \restr{\prnt{\_}}{F'}, \agrp', \acrd')$,  where $F' = \gchld{f} \cup \{f\}$, $\agrp' = \agrp \cap 2^{F'}$, and $\acrd' = \acrd|_{F'\cup \agrp'}$. 
 \qed
\end{defin}

\begin{figure}
\centering
       \includegraphics[scale=0.5
                 ]
                 {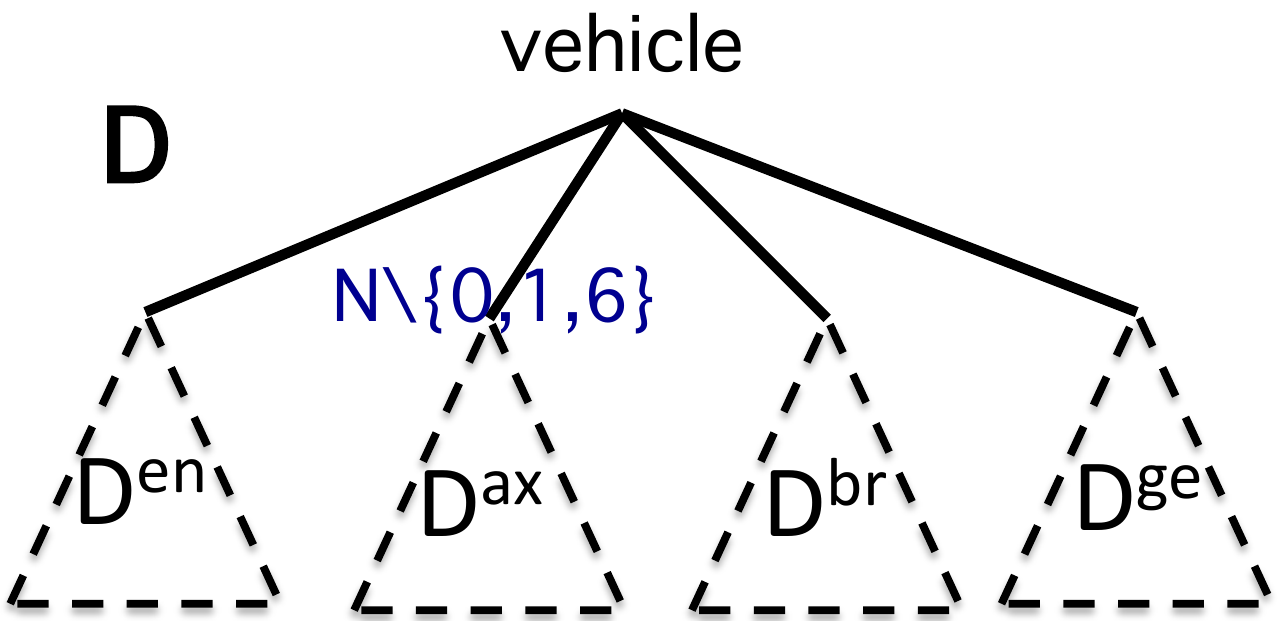}
\caption{The representation of \ancd\ in \figref{fig:CFD-intro}  in terms of induced diagrams} 
\label{fig:exCFD-running-induced} 
\end{figure}
In the hierarchical theory of a given CFD, the multiplicity domain of a solitary feature is considered as a multiplicity constraint on its corresponding induced diagram. Looking at \figref{fig:exCFD-running-induced}, which represents the CFD \ancd\ in \figref{fig:CFD-intro} in terms of induced diagrams: The root feature \veh\ has four children labeled by $\inducN{\ancd}{\en}$ (diagram induced by \eng),  $\inducN{\ancd}{\ax}$ (diagram induced by \axle),  $\inducN{\ancd}{\br}$ (diagram induced by \brake), and $\inducN{\ancd}{\ger}$ (diagram induced by \gear) with multiplicity domains $\{1\}$, $\nat\setminus \{0,1,6\}$, $\{1\}$, and $\{1\}$, respectively.  
In this way, a hierarchical product of \ancd\ is considered as a multiset $\lceil \veh, h_\en, h_\ax^{c_1}, h_\br, h_\ger \rceil$, where $h_\en$, $h_\ax$, $h_\br$, and $h_\ger$ are a hierarchical product of $\inducN{\ancd}{\en}$, a hierarchical product of $\inducN{\ancd}{\ax}$, a hierarchical product of $\inducN{\ancd}{\br}$, and a hierarchical product of $\inducN{\ancd}{\ger}$, respectively, and $c_1\in \nat \setminus \{1, 6\}$ (the multiplicity domain of \axle). Note that the multiplicities of $h_\en$, $h_\br$, and $h_\ger$ are always 1, as the multiplicity domains of \eng, \brake, and \gear\ are all $\{1\}$.   
\begin{figure}[h]
\centering
    \includegraphics[scale=0.5]{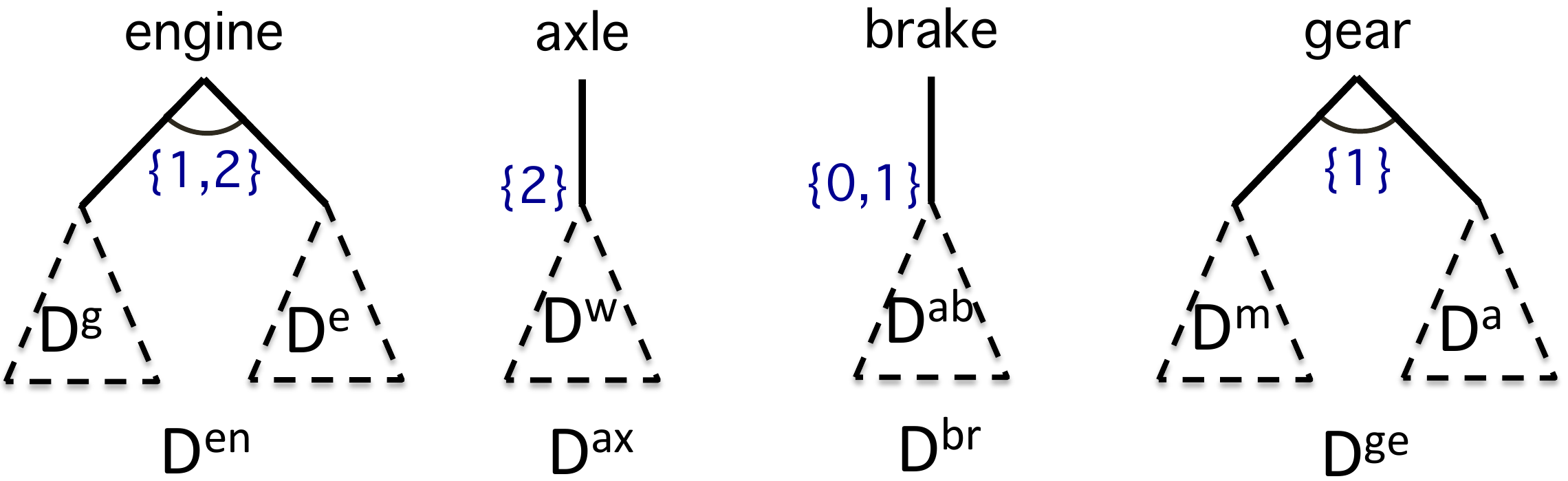}
\caption{CFDs in terms of induced diagrams}
\label{fig:cfd-induc-others}
\end{figure}

Now, consider $\inducN{\ancd}{\en}$ (diagram induced by \eng) shown in \figref{fig:cfd-induc-others}. The feature \eng\ has 2 subfeatures \gas\ and \elec\ which are grouped together with multiplicity $\{1,2\}$.  To distinguish between grouped and solitary features, we introduce a notion, called {\em  hierarchical products associated with groups}: 

Suppose that we choose both $\gas$ and $\elec$ from the group $\xxx{G}$ in our configuration (the multiplicity domain of \xxx{G} allows us to do so). A corresponding hierarchical product associated with the group would be a multiset $\lceil h_\g, h_\el^{c_3} \rceil$, where $h_\g$ and $h_\el$ are, respectively, a hierarchical product of $\inducN{\ancd}{\g}$ (diagram induced by \gas) and a hierarchical product of $\inducN{\ancd}{\el}$ (diagram induced by \elec). The multiplicity of $h_\g$ is always $1$, as the multiplicity domains of $\gas$ is the singleton $\{1\}$. The multiplicity of $h_\el$, $c_3$, must be in $\{1,2\}$ (the multiplicity domain of \elec). Since $\inducN{\ancd}{\g}$ and $\inducN{\ancd}{\el}$ are singleton trees, $h_\g$ and $h_\el$ would be always $\lceil \gas \rceil$ and $\lceil \elec \rceil$, respectively.  Another example: let us now choose only \gas\ from the group (note that, due to the multiplicity domain of \xxx{G}, this configuration is valid). Then, the corresponding hierarchical product would be a multiset $\lceil  h_\g \rceil$. The notion hierarchical products associated with groups allows us to ``explicitly'' distinguish between grouped and solitary features.

This way, a hierarchical product of $\inducN{\ancd}{\en}$ would be a multiset $\lceil \eng, h_G \rceil$, where $h_G$ is a hierarchical product associated with the group $\xxx{G} = \{\gas, \elec\}$.

A hierarchical product of $\inducN{\ancd}{\ax}$ (diagram induced by \axle) is a multiset $\lceil \axle, h_\w^{2} \rceil$, where $h_\w$ is a hierarchical product of $\inducN{\ancd}{\w}$  (the diagram induced by \wheel) -- see \figref{fig:cfd-induc-others}. Since $\inducN{\ancd}{\w}$  is a singleton CFD, $h_\w$ is always equal to $\lceil \wheel \rceil$.

A hierarchical product of $\inducN{\ancd}{\br}$ (diagram induced by \brake) would be a multiset $\lceil \brake, h_\ab^{c_2} \rceil$, where $h_\ab$ is a hierarchical product of $\inducN{\ancd}{\ab}$ (diagram induced by \abs) and $c_2 \in \{0,1\}$ -- see \figref{fig:cfd-induc-others}. Note that $h_\ab$ is always $\lceil \abs \rceil$, as $\inducN{\ancd}{\ab}$ is  a singleton CFD. 


Now, consider $\inducN{\ancd}{\ger}$ (diagram induced by \gear) in \figref{fig:cfd-induc-others}. The feature $\gear$ has 2 subfeatures $\manu$ and $\auto$ which are grouped together with multiplicity $\{1\}$. A hierarchical product of $\inducN{\ancd}{\ger}$ would be a multiset $\lceil \gear, h_{G'} \rceil$, where $h_{G'}$ is a hierarchical product associated with the group $\xxx{G'} = \{\manu, \auto\}$.


According to discussion above, a hierarchical product of the CFD in \figref{fig:CFD-intro} would be a multiset 
$$\lceil \ve, \lceil \en, \lceil \lceil \g \rceil^{g_1}, \lceil \el \rceil^{c_3\times g_2} \rceil \rceil,  \lceil \ax, \lceil \w \rceil^2 \rceil^{c_1}, \lceil \br, \lceil \ab \rceil^{c_2} \rceil  , \lceil \ger, \lceil \lceil \man \rceil^{g_3}, \lceil \aut \rceil^{g_4} \rceil \rceil \rceil,$$ 
where \ve, \en, \g, \el, \ax, \w, \br, \ab, \ger, \man, and \aut stand for \veh, \eng, \gas, \elec, \axle, \wheel, \brake,  \abs, \gear, \manu, and \auto, respectively, and $g_{1-4} \in \{0,1\}$,  $1\leq g_1 + g_2 \leq 2$, $g_3 + g_4 = 1$, $c_1 \in \nat\setminus\{0,1,6\}$, $c_2 \in \{0,1\}$, and $c_3 \in \{1,2\}$.  The conditions $1 \leq g_1 + g_2 \leq 2$ and $g_3 + g_4 = 1$ ensure that the  multiplicity constraints $\{1,2\}$ and $\{1\}$ on the groups $\{\gas, \elec\}$ and $\{\manu, \auto\}$, respectively, are satisfied.\footnote{ Note that an element in a multiset with multiplicity 0 means that the element does not belong to the domain of the multiset, \eg, $\lceil a, \lceil b \rceil^0 \rceil = \lceil a \rceil$.}   

According to above,  the following multisets are valid hierarchical products of the CFD in \figref{fig:CFD-intro}, where \ve, \en, \g, \el, \ax, \w, \br, \ab, \ger, \man, and \aut stand for \veh, \eng, \gas, \elec, \axle, \wheel, \brake,  \abs, \gear, \manu, and \auto, respectively:
\label{pp:hierex}
 $$h_1 = \lceil \ve, \lceil \en, \lceil \lceil \g \rceil \rceil \rceil, \lceil \ax, \lceil \w \rceil^2 \rceil^{3}, \lceil \br \rceil, \lceil \ger, \lceil \lceil \man \rceil \rceil \rceil  \rceil.$$ 
$$h_2 = \lceil \ve, \lceil \en, \lceil \lceil \el \rceil \rceil \rceil, \lceil \ax, \lceil \w \rceil^2 \rceil^{3}, \lceil \br \rceil, \lceil \ger, \lceil \lceil \man \rceil \rceil \rceil  \rceil.$$  
$$h_3 = \lceil \ve, \lceil \en, \lceil \lceil \el \rceil^2 \rceil \rceil, \lceil \ax, \lceil \w \rceil^2 \rceil^{3}, \lceil \br \rceil, \lceil \ger, \lceil \lceil \man \rceil \rceil \rceil  \rceil.$$
$$h_4 = \lceil \ve, \lceil \en, \lceil \lceil \g \rceil, \lceil \el \rceil \rceil \rceil, \lceil \ax, \lceil \w \rceil^2 \rceil^{3}, \lceil \br \rceil, \lceil \ger, \lceil \lceil \man \rceil \rceil \rceil  \rceil.$$
$$h_5 = \lceil \ve, \lceil \en, \lceil \lceil \g \rceil, \lceil \el \rceil^2 \rceil \rceil, \lceil \ax, \lceil \w \rceil^2 \rceil^{3}, \lceil \br \rceil, \lceil \ger, \lceil \lceil \man \rceil \rceil \rceil  \rceil.$$
$$h_6 = \lceil \ve, \lceil \en, \lceil \lceil \g \rceil \rceil \rceil, \lceil \ax, \lceil \w \rceil^2 \rceil^{3}, \lceil \br \rceil, \lceil \ger, \lceil \lceil \aut \rceil \rceil \rceil  \rceil.$$
$$h_7 = \lceil \ve, \lceil \en, \lceil \lceil \el \rceil \rceil \rceil, \lceil \ax, \lceil \w \rceil^2 \rceil^{3}, \lceil \br \rceil, \lceil \ger, \lceil \lceil \aut \rceil \rceil \rceil  \rceil.$$ 
$$h_8 = \lceil \ve, \lceil \en, \lceil \lceil \el \rceil^2 \rceil \rceil, \lceil \ax, \lceil \w \rceil^2 \rceil^{3}, \lceil \br \rceil, \lceil \ger, \lceil \lceil \aut \rceil \rceil \rceil  \rceil.$$
$$h_9 = \lceil \ve, \lceil \en, \lceil \lceil \g \rceil, \lceil \el \rceil \rceil \rceil, \lceil \ax, \lceil \w \rceil^2 \rceil^{3}, \lceil \br \rceil, \lceil \ger, \lceil \lceil \aut \rceil \rceil \rceil  \rceil.$$
$$h_{10} = \lceil \ve, \lceil \en, \lceil \lceil \g \rceil, \lceil \el \rceil^2 \rceil \rceil, \lceil \ax, \lceil \w \rceil^2 \rceil^{3}, \lceil \br \rceil, \lceil \ger, \lceil \lceil \aut \rceil \rceil \rceil  \rceil.$$
$$h_{11} = \lceil \ve, \lceil \en, \lceil \lceil \g \rceil \rceil \rceil, \lceil \ax, \lceil \w \rceil^2 \rceil^{3}, \lceil \br , \lceil \ab \rceil \rceil, \lceil \ger, \lceil \lceil \man \rceil \rceil \rceil  \rceil.$$
$$h_{12} = \lceil \ve, \lceil \en, \lceil \lceil \el \rceil \rceil \rceil, \lceil \ax, \lceil \w \rceil^2 \rceil^{3}, \lceil \br, \lceil \ab \rceil \rceil, \lceil \ger, \lceil \lceil \man \rceil \rceil \rceil  \rceil.$$
$$h_{13} = \lceil \ve, \lceil \en, \lceil \lceil \el \rceil^2 \rceil \rceil, \lceil \ax, \lceil \w \rceil^2 \rceil^{3}, \lceil \br, \lceil \ab \rceil \rceil, \lceil \ger, \lceil \lceil \man \rceil \rceil \rceil  \rceil.$$
$$h_{14} = \lceil \ve, \lceil \en, \lceil \lceil \g \rceil, \lceil \el \rceil \rceil \rceil, \lceil \ax, \lceil \w \rceil^2 \rceil^{3}, \lceil \br, \lceil \ab \rceil \rceil, \lceil \ger, \lceil \lceil \man \rceil \rceil \rceil  \rceil.$$
$$h_{15} = \lceil \ve, \lceil \en, \lceil \lceil \g \rceil, \lceil \el \rceil^2 \rceil \rceil, \lceil \ax, \lceil \w \rceil^2 \rceil^{3}, \lceil \br, \lceil \ab \rceil \rceil, \lceil \ger, \lceil \lceil \man \rceil \rceil \rceil  \rceil.$$
$$h_{16} = \lceil \ve, \lceil \en, \lceil \lceil \g \rceil \rceil \rceil, \lceil \ax, \lceil \w \rceil^2 \rceil^{3}, \lceil \br, \lceil \ab \rceil \rceil, \lceil \ger, \lceil \lceil \aut \rceil \rceil \rceil  \rceil.$$
$$h_{17} = \lceil \ve, \lceil \en, \lceil \lceil \el \rceil \rceil \rceil, \lceil \ax, \lceil \w \rceil^2 \rceil^{3}, \lceil \br, \lceil \ab \rceil \rceil, \lceil \ger, \lceil \lceil \aut \rceil \rceil \rceil  \rceil.$$
$$h_{18} = \lceil \ve, \lceil \en, \lceil \lceil \el \rceil^2 \rceil \rceil, \lceil \ax, \lceil \w \rceil^2 \rceil^{3}, \lceil \br, \lceil \ab \rceil \rceil, \lceil \ger, \lceil \lceil \aut \rceil \rceil \rceil  \rceil.$$
$$h_{19} = \lceil \ve, \lceil \en, \lceil \lceil \g \rceil, \lceil \el \rceil \rceil \rceil, \lceil \ax, \lceil \w \rceil^2 \rceil^{3}, \lceil \br, \lceil \ab \rceil \rceil, \lceil \ger, \lceil \lceil \aut \rceil \rceil \rceil  \rceil.$$
$$h_{20} = \lceil \ve, \lceil \en, \lceil \lceil \g \rceil, \lceil \el \rceil^2 \rceil \rceil, \lceil \ax, \lceil \w \rceil^2 \rceil^{3}, \lceil \br, \lceil \ab \rceil \rceil, \lceil \ger, \lceil \lceil \aut \rceil \rceil \rceil  \rceil.$$

Consider again the CFDs $\ancd_1$ and $\ancd_2$ in \figref{fig:2-flat-equi}. Unlike their flat theories, their hierarchical theories capture the differences:  
$\ancd_1$ contains the two hierarchical products $\lceil\xxxf, \lceil\xf{2}, \lceil\xf{3}\rceil\rceil\rceil$ and $\lceil \xxxf, \lceil\xf{2}, \lceil \xf{3}\rceil\rceil^2\rceil$, while $\ancd_2$ contains $\lceil \xxxf, \lceil \xf{3}, \lceil\xf{2}\rceil \rceil \rceil$ and $\lceil \xxxf, \lceil \xf{3}, \lceil \xf{2} \rceil \rceil^2 \rceil$ as its hierarchical products. 

In the rest of this section, we formalize the hierarchical multiset theories of CFDs and prove some results. We first define a hierarchy of multisets over a set of urelements\footnote{~ An urlement is an object, which may be an element of a set or multiset, but it is not a set or multiset.} whose first class is the set of finite multisets over the features and other classes are defined as the set of all finite multisets built over the union of the previous classes and the set itself. This hierarchy  will be a fundamental basis for formalizing the hierarchical theories of CFDs. Since the set of features in a CFD is a finite set, we will always deal with finite multisets. Let $\multisets{A}$ denote the class of all finite multisets over $A$. 

\begin{defin}[A Hierarchy of Finite Multisets] \label{def:hierarchy-multiset}
For every nonempty set of urelements $A$, we define a {\em hierarchy} $\multhier(A)$ of multisets as follows: 
$$
\multhier_1(A) = \multisets{A},  \ldots \multhier_{n+1} = \multisets{A \cup \bigcup_{0\leq i\leq n} \multhier_i}, \ldots. 
$$
$$
 \multhier(A) = \bigcup_{i\geq 1} \multhier_i(A)
$$
The {\em rank} of a multiset $m\in \multhier(A)$, denoted by $\rankm{m}$, is equal to the least number $n$ such that $m\in \multhier_n(A)$. Any multiset with rank 1 is called a {\em flat multiset} over $A$.  
\qed
\end{defin}
As an example, consider  the multisets $m_1 = \lceil a^3, b^3 \rceil$, $m_2 = \lceil a^2, \lceil a^2, b^3 \rceil, \lceil b \rceil^4 \rceil$,  and $m_3 = \lceil a^{10}, \lceil a^2, b^3 \rceil^3, \lceil b \rceil^4, \lceil \lceil a\rceil\rceil \rceil$ in $\multhier(\{a, b\})$. Their ranks are as follows: $\rankm{m_1} = 1$, $\rankm{m_2} = 2$, and $\rankm{m_3} = 3$.
%
%
%
%

Now, we are at the point where we can formalize hierarchical products of CFDs. Consider a CFD $\ancd \in \cfdfam(F)$. Suppose that its root has $n$ solitary subfeatures $s_1, \ldots, s_n$ and $k$ groups $G_1,  \ldots, G_k$. According to our description of hierarchical products, any multiset $m\in \multhier(F)$ is a hierarchical product of \ancd\ if its domain consists of (i) $r$ with 1 occurrence, (ii) a  hierarchical product of $\inducN{\ancd}{s_i}$ (diagram induced by $s_i$) with a multiplicity  $c_i\in \acrd(s_i)$ for any $1\leq i \leq n$, (iii) a hierarchical product associated with $G_j$  with multiplicity 1 for any $1\leq j \leq k$. Hierarchical products of a given CFD are formalized as follows: 

\begin{defin}[Hierarchical Products] \label{def:hier-products}
Given a CFD $\ancd = (F, r, \prnt{\_}, \agrp, \acrd)$, the set of \ancd's {\em hierarchical products}, denoted by $\products(\ancd)$, is defined as follows: 
For any  multiset $m\in \multhier(F)$,  $m\in \products(\ancd)$ if it satisfies the following conditions: 

(i)  $m(r) = 1$.

(ii) $\forall f\in \asol \cap \chld{r}, \exists c\in \acrd(f), \exists n\in \products(\inducN{\ancd}{f}):  m(n) = c$. 

(iii) $\forall G \in \agrp \cap 2^{\chld{r}}, \exists n \in \products(\ancd, G): m(n) = 1$. 
 
 ~~~~~(see \defref{def:g-products} for the definition of  $\products(\ancd, G)$)
\\
$\products(\ancd)$ is  called the {\em hierarchical theory} of \ancd.
\qed
\end{defin}

\begin{defin}[Hierarchical Products Associated with Groups]\label{def:g-products}
~\\
Let $\ancd = (F, r, \_^\uparrow, \agrp, \acrd)$ be a CFD and $G = \{f_1, f_2, \ldots, f_k\}\in \agrp$ for some $k$. A {\em hierarchical product associated with $G$} is a multiset $m\in \multhier(F)$ such that for all $1\leq i \leq k$, there exist $c\in \acrd(G)$,  $c_i \in \acrd(f_i)$, $g_i \in \{0,1\}$, $m_i\in \products(\inducN{\ancd}{f_i})$, and

(i) $\dom{m} = \{m_i: g_i = 1\}$,

(ii) $\forall 1\leq i \leq k: m(m_i) = c_i\times g_i$,

(iii) $g_1 + \ldots + g_k = c$.
\\
The set of hierarchical products associated with $G$ is denoted by $\products(\ancd, G)$.
 \qed
\end{defin}
The following theorem shows that, unlike the flat theory, the hierarchical theory of a given CFD captures all information of the CFD. 
\begin{theo} \label{th:unique}
Given two CFDs \ancd\ and $\ancd'$, $(\products(\ancd) = \products(\ancd')) \biimplies (\ancd = \ancd')$. \qed
\end{theo}
The cardinalities of the hierarchical and flat theories of a given CFD are the same, \ie, there is a bijection between the set of hierarchical products and the set of flat products for a ``fixed'' CFD. We will show this in \theoref{th:hier-flat}. Before getting to this formally, we first need to define some notions.  

The domain of a multiset with rank greater than 1 includes some multisets.  For example consider the multiset $m = \lceil a, b, \lceil c, \lceil d, e \rceil \rceil \rceil \in \multhier_3(\{a,b,c,d,e\})$. The domain of this multiset includes the multiset $i_1 = \lceil c, \lceil d, e \rceil \rceil$. The domain of $i_1$ itself includes the multiset $i_2 = \lceil d, e \rceil$ whose domain is a set of urelements. We call $i_1$ and $i_2$ the {\em multiset ingredients} of $m$. 
\begin{defin}[Multiset Ingredients of Multisets] \label{def:mult-ing}
Given a multiset $m\in \multhier(A)$ for some $A$, the {\em multiset ingredients} of $m$, denoted by  $\multing{m}$, is the smallest set of multisets in $\multhier(A)$ such that

(i) $\{n\in dom(m): \rankm{n}\geq 1\} \subseteq \multing{m}$,

(ii) $\forall n\in \multing{m}: \multing{n} \subset \multing{m}$.
\\
The multiplicity of a multiset $n\in \multing{m}$ in $m$ is denoted by $\#_m(n)$.\qed
\end{defin}
The following definition formalizes a notion called the {\em flat multiplicity} of an urelement in a multiset.  An illustrating example follows the definition. 
\begin{defin}[Flat Multiplicities and Flattening] \label{def:flatten}
Let $m\in \multhier(A)$ for a set $A$ of urelements. The {\em flat multiplicity} of an element  is defined by a function  $\#_{m, A}: A\rarr \nat$ as  $\#_{m, A}(a) = m(a) + \sum_{e\in \multing{m}} \#_{m, A}(e)$. 

We define a function $\xxx{flat}_A: \multhier(A) \rarr \multhier_1(A)$, which maps a given multiset $m \in \multhier(A)$ to a flat multiset as follows. For any $m\in \multhier(A): \pure{A}{m}(a) = \#_{m, A}(a)$.  We say that $\pure{A}{m}$ {\em flattens} $m$.
\qed
\end{defin}
Consider  $m = \lceil a^2, b^2,  \lceil a^8,  \lceil a^5, b^3 \rceil^3\rceil\rceil$. The flat multiplicities of $a$  and $b$  are $25$ and $11$, respectively. Thus, $\pure{\{a,b\}}{m} = \lceil a^{25}, b^{11} \rceil$.

Consider the following hierarchical products of the CFD in \figref{fig:CFD-intro}, where \ve, \en, \g, \el, \ax, \w, \br, \ab, \ger, \man, and \aut stand for \veh, \eng, \gas, \elec, \axle, \wheel, \brake,  \abs, \gear, \manu, and \auto, respectively: 

 --- $h_1 = \lceil \ve, \lceil \en, \lceil \lceil \g \rceil \rceil \rceil, \lceil \ax, \lceil \w \rceil^2 \rceil^{3}, \lceil \br \rceil, \lceil \ger, \lceil \lceil \man \rceil \rceil \rceil  \rceil$.  

--- $h_9 = \lceil \ve, \lceil \en, \lceil \lceil \g \rceil, \lceil \el \rceil \rceil \rceil, \lceil \ax, \lceil \w \rceil^2 \rceil^{3}, \lceil \br \rceil, \lceil \ger, \lceil \lceil \aut \rceil \rceil \rceil  \rceil$.

--- $h_{20} = \lceil \ve, \lceil \en, \lceil \lceil \g \rceil, \lceil \el \rceil^2 \rceil \rceil, \lceil \ax, \lceil \w \rceil^2 \rceil^{3}, \lceil \br, \lceil \ab \rceil \rceil, \lceil \ger, \lceil \lceil \aut \rceil \rceil \rceil  \rceil$.   
 
 Flattening them,  we obtain, respectively, the following flat products of the CFD: 
 
 -- $m_1 =  \lceil \veh, \gear, \brake, \eng, \gas, \axle^3, \wheel^6, \manu  \rceil$. 

 - $m_9 =  \lceil \veh, \gear, \brake, \eng, \gas, \elec, \axle^3, \wheel^6, \auto  \rceil$. 
 
 -- $m_{20} =   \lceil \veh, \gear, \brake, \eng, \gas, \elec^2, \axle^3, \wheel^6, \auto, \abs  \rceil$. 
 
 Indeed, the hierarchical products  $h_{1-20}$ of the CFD in \figref{fig:CFD-intro} (see page \pageref{pp:hierex}) are hierarchal versions of the flat products $m_{1-20}$ of the CFD (see page \pageref{pp:flatp}). 

The following theorem shows that the restriction of the flattening function to the domain of the hierarchical theory of a given CFD provides a bijection between the two multiset theories of the CFD. 

\begin{theo}\label{th:hier-flat}
For any CFD $\ancd\in \cfdfam(F)$, the function $\restr{\xxx{flat}_F}{\products(\ancd)}$, \ie, the restriction of $\xxx{flat}_F$ to the subdomain $\products(\ancd)$, provides a bijection from $\products(\ancd)$ to $\pproducts(\ancd)$. \qed
\end{theo}
{\bf Remark.} 
CFDs subsume BFDs. Therefore, the work presented here for CFDs can be applied on BFDs too.   In this sense, a flat product of a BFD is a set of features (in other words, a multiset of features in which the multiplicities of the  elements in the multiset are all 1), and a hierarchical product of the BFD is a hierarchical set over features (in other words, a hierarchical multiset on which the multiplicities of any  element of its domain as well as the multiplicity of any multiset ingredients are all 1). 

\section{Characterization of  Hierarchical Products
} \label{sec:tree-mult}
In this section, we characterize the domain of multisets that can be hierarchical products of some CFDs. We first define a notion called {\em tree-like multisets} over a given set $F$. The set of tree-like multisets over $F$ is denoted by $\treemulthier(F)$, which is the restriction of $\multhier(F)$ to tree-like multisets. We then show that a multiset $m \in \multhier(F)$ can be a hierarchical product of some CFDs in $\cfdfam(F)$ if and only if it is a tree-like multiset over $F$. A definition of tree-like multisets is as follows (illustrating examples follow the formal definitions).  


\begin{defin}[{\bf Tree-like Multisets}]  \label{def:tree-mult}
Given a set $F$, the set of {\em tree-like multisets} over $F$, denoted by $\treemulthier(F)$, is inductively defined  as follows:


(i)  $\lceil f \rceil \in \treemulthier(F)$, $\forall f\in F$. 


(ii) $t_1\uplus \lceil t_2^n \rceil \in \treemulthier(F)$,  $\forall t_1, t_2\in \treemulthier(F), \forall n\in \nat$, if 

~~~~ $\dom{\pure{F}{t_1}} \cap \dom{\pure{F}{t_2}} = \varnothing$.



(iii) $t \uplus \lceil \lceil ~ t_1^{n_1}, \ldots, t_k^{n_k}~ \rceil\rceil \in \treemulthier(F)$, $\forall t, t_1, \dots t_k \in \treemulthier(F), \forall n_1, \ldots, n_k \in \nat$, if 

~~~~ $\forall 1\leq i \leq k: \dom{\pure{F}{t}} \cap \dom{\pure{F}{t_i}} = \varnothing$, and 

~~~~ $\forall 1\leq i, j \leq k: (i\neq j) \implies (\dom{\pure{F}{t_i}} \cap \dom{\pure{F}{t_j}} = \varnothing)$.
%
%
%
%
%
\qed
\end{defin}
For example, the following multisets in $\multhier(\{a, b, c\})$ are tree-like multisets: 

-- $t_1  = \lceil a \rceil$, $t_2 = \lceil b \rceil$, and $t_3 = \lceil c \rceil$,

-- $t_4 = t_1 \uplus \lceil t_2^6 \rceil = \lceil a, \lceil b\rceil^6\rceil$, $t_5 = t_3 \uplus \lceil t_4 \rceil = \lceil c, \lceil a, \lceil b\rceil \rceil^6\rceil$, 

-- $t_6 = t_1 \uplus \lceil \lceil t_2^2, t_3 \rceil \rceil = \lceil a,  \lceil\lceil b \rceil^2, \lceil c \rceil \rceil \rceil$.  

The following multisets are not valid tree-like multisets:

-- $n_1 = \lceil a, b \rceil$, 

-- $n_2 = \lceil a^3, \lceil b\rceil^6\rceil$, 

-- $n_3 =  \lceil a,  \lceil\lceil b, c \rceil^2 \rceil \rceil$.

Restriction of $\multhier(A)$ to  tree-like multisets results in a hierarchy of tree-like multisets. Let us denote this hierarchy and its classes  by $\treemulthier(A)$ and $\treemulthier_i(A)$ ($i \geq 1$), respectively. According to \defref{def:tree-mult}, $\treemulthier_1(A) = \{\lceil a \rceil: a\in A\}$ and  $\treemulthier(A) = \bigcup_{i} \treemulthier_i(A)$.  Note that $\treemulthier(A)$ is not closed under additive union and multiset minus.

\begin{defin} [{\bf Groups of Tree-like Multisets}] \label{def:group-treemult}
Given a set $F$, tree-like multisets $t_1, \dots t_k \in \treemulthier(F)$, and integers $n_1, \ldots, n_k \in \nat$, the multiset $\lceil\lceil t_1^{n_1}, \ldots,$  $t_k^{n_k} \rceil\rceil \in \multhier(F)$ is called a {\em group of tree-like multisets} over $F$ if 
%
%

$\forall 1\leq i, j \leq k: (i\neq j) \implies (\dom{\pure{F}{t_i}} \cap \dom{\pure{F}{t_j}} = \varnothing).$

Any element of the domain of a group of tree-like multisets is called a {\em grouped tree-like multiset}. \qed  
\end{defin}
The multiset $\lceil\lceil b \rceil^2, \lceil c, [a]^3 \rceil \rceil$ is an example of a group of tree-like multisets over $\{a, b, c\}$. 

As noticed, the domain of a tree-like multiset includes a unique urelement with multiplicity 1. We call this element the {\em root} of the tree-like multiset, formalized in the following definition. 
%
%
\begin{defin} [{\bf Roots of Tree-like Multisets}]
Given a set $F$, we define a function $root: \treemulthier(F) \rarr F$, as follows:

(i) $\rootm{\lceil f \rceil} = f$, for any $f\in F$.

(ii) $\rootm{t_1\uplus \lceil t_2^n \rceil} = \rootm{t_1}$ for any $t_1, t_2\in \treemulthier(F), n\in \nat$ satisfying the conditions in \defref{def:tree-mult}(ii).

(iii) $\rootm{t \uplus \lceil ~t_1^{n_1}, \ldots, t_k^{n_k}~\rceil} = \rootm{t}$ for any $t, t_1, \dots t_k \in \treemulthier(F)$ and $n_1, \ldots, n_k \in \nat$ satisfying the conditions in \defref{def:tree-mult}(iii)
%
\qed
\end{defin}
Note that any multiset ingredient of a tree-like multiset is either a tree-like multiset or a group of tree-like multisets.  
As an example, the multiset $ t = \lceil  a, \lceil b \rceil^5, \lceil c, \lceil d \rceil^3, \lceil \lceil e \rceil,$ $\lceil f \rceil \rceil\rceil^2 \rceil$ is a tree-like multiset over the set $\{a, b, c, d, e, f\}$: $\rootm{t} = a$; the elements $t_1 = \lceil b \rceil, t_2 = \lceil c, \lceil d \rceil^3, \lceil \lceil e \rceil,$ $\lceil f\rceil\rceil\rceil \in \dom{t}$ are both tree-like multisets with $\rootm{t_1} = b$ and $\rootm{t_2} = c$, respectively; the element $\lceil\lceil e\rceil, \lceil f \rceil \rceil \in \dom{t_2}$ is a group of  tree-like multisets. 

 The following theorem shows that a hierarchical product of a CFD is always a tree-like multiset.
\begin{theo} \label{th:hierp-2-treemult}
Any hierarchical product of a given CFD over a set of features $F$ is a tree-like multiset over $F$. \qed
\end{theo}

For example, (it is easy to see that) the hierarchical products  $h_{1-20}$ of the CFD in \figref{fig:CFD-intro} (see page \pageref{pp:hierex}) are all tree-like multisets. 

The rest of the section is devoted to showing that  any tree-like multiset is a hierarchical product of some CFDs. 
%
%
%
%
We show how to extract a  CFD from a given tree-like multiset. This is done step by step through the following definitions. Definitions \ref{def:tree-treemult}, \ref{def:groups-treemult}, and \ref{def:card-treemult} show, respectively, how to extract the tree, groups, and multiplcities  from a given tree-like multiset. 

We first  define the notion of a {\em tree-like multisest  induced} by an element:
\begin{defin}[{\bf Tree-like Multiset Induced  by Elements}] \label{def:induced-tree}
For a given tree-like multiset $t$ over a set $F$, the {\em tree-like multiset induced} by $f$, denoted by $\inducN{t}{f}$, is the multiset ingredient of $t$ whose root is $f$. \qed
\end{defin}
\begin{remar}
According to \defref{def:tree-mult}, the following statement follows obviously: 
Let $t\in \treemulthier(F)$ for a set  $F$. For any $f\in \dom{\pure{F}{t}}$, there is a unique multiset ingredient of $t$ whose root is  $f$. This uniqueness makes \defref{def:induced-tree}  well-formed. \qed
\end{remar}
For an example, consider the tree-like multiset $ t = \lceil a, \lceil b\rceil^5, \lceil c, \lceil d\rceil^3, \lceil \lceil e\rceil, \lceil f\rceil\rceil\rceil^2\rceil$ over  the set $\{a, b, c, d, e, f\}$. Then, we would have: 
~$\inducN{t}{a} = t$, 
~$\inducN{t}{b} = \lceil b\rceil$, 
 ~$\inducN{t}{c} = \lceil c, \lceil d\rceil^3, \lceil\lceil e\rceil, \lceil f\rceil\rceil\rceil$, 
 ~$\inducN{t}{d} = \lceil d \rceil$, 
~ $\inducN{t}{e} = \lceil e \rceil$, and 
~ $\inducN{t}{f} = \lceil f \rceil$.  

The following definition introduces the notion of {\em parents} in  tree-like multisets. 
\begin{defin}[{\bf Parents of Elements in Tree-like Multisets}] \label{def:treemult-parent}
For a given tree-like multiset $t$ over a set $F$ and $f\in \dom{\pure{F}{t}} \setminus \{\rootm{t}\}$, the {\em parent of $f$}, denoted by $\prntmult{t}{f}$,  is an element in $\dom{\pure{F}{t}}$ such that 


(i) if $\inducN{t}{f}$ is a grouped tree-like multiset under a group $g$ of multisets, then $g$ is in the domain of the  tree-like multiset induced by $\prntmult{t}{f}$, \ie,   $g\in \dom{\inducN{t}{\prntmult{t}{f}}}$.


(ii) if $\inducN{t}{f}$ is a tree-like multiset, then it is in the domain of the  tree-like multiset induced by $\prntmult{t}{f}$, \ie,   $\inducN{t}{f} \in \dom{\inducN{t}{\prntmult{t}{f}}}$.
%
 \qed
\end{defin}
\begin{remar}
According to \defref{def:tree-mult}, the following statement follows obviously:  
Consider a tree-like multiset $t\in \treemulthier(F)$ for a set $F$. For any $f\in \dom{\pure{F}{t}} \setminus \{\rootm{t}\}$, there exists a unique element in $\dom{\pure{F}{t}}$ satisfying (i) and (ii) in \defref{def:treemult-parent}.  This makes \defref{def:treemult-parent} well-formed. 
Therefore, $\prntmult{t}{\_}$ is indeed a function from  $\dom{\pure{F}{t}} \setminus \{\rootm{t}\}$ to $\dom{\pure{F}{t}}$. \qed
\end{remar} 

 For an example, consider the tree-like multiset $t = \lceil a, \lceil b\rceil^5, \lceil c, \lceil d\rceil^3, \lceil\lceil e\rceil, \lceil f\rceil\rceil\rceil^2\rceil$ over  the set $\{a, b, c, d, e, f\}$. We would have: ~$\prntmult{t}{b} = \prntmult{t}{c} = a$ and ~  $\prntmult{t}{d} = \prntmult{t}{e} = \prntmult{t}{f} = c$. 

Now we can  see that any tree-like multiset represents a unique tree of the elements of its corresponding flat multiset. This tree is extracted using the parents of elements. The following definition shows how to do so. 
\begin{defin}[{\bf Trees Associated with Tree-like Multisets}] \label{def:tree-treemult}
Let  $t \in \treemulthier(F)$.  The {\em tree associated} with $t$, denoted by $\treem{t}$, is defined as follows: $\treem{t} = (N_t, r_t, {\_}^{\uparrow_t})$, where $N_t = \dom{\pure{F}{t}}$, $r_t = \rootm{t}$, and  $\prntmult{t}{\_}: N_t\setminus \{r_t\} \rarr N_t$ is a function defined in \defref{def:treemult-parent}.  
%
\qed
\end{defin}

\begin{figure}[h]
\centering
    \includegraphics[scale=0.45]{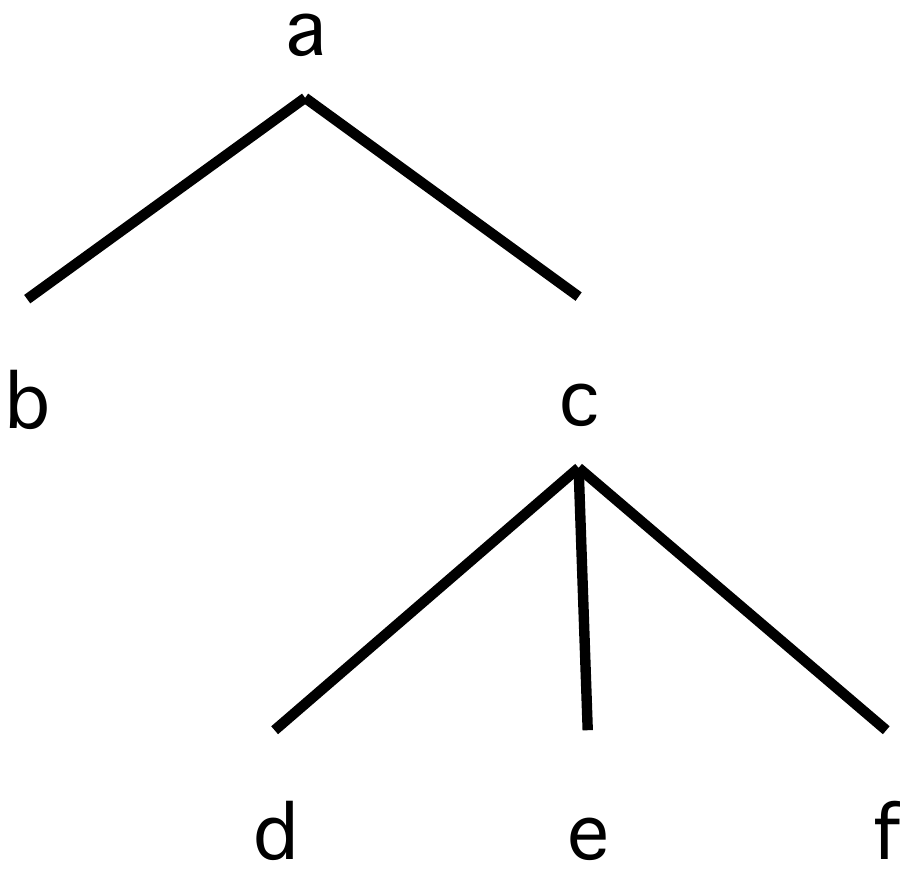}
\caption{Trees associated with tree-like multisets: example}
\label{fig:tree-treemult-ex}
\end{figure}

For an example, the tree associated with  $t = \lceil a, \lceil b\rceil^5, \lceil c, \lceil d\rceil^3, \lceil \lceil e \rceil, \lceil f \rceil \rceil\rceil^2\rceil \in \treemulthier(\{a, b, c, d, e, f\})$ is represented in \figref{fig:tree-treemult-ex}: The root of $t$, \ie, $a$, is the root of the tree. There are two elements, $b$ and $c$, whose parents are $a$.  $b$ is a leaf in the tree, as there is no element whose parent is $b$. There are three elements $d, e$, and $f$ whose parents are $c$ ($e$ and $f$ are grouped tree-like multisets and their corresponding group $\lceil \lceil e\rceil, \lceil f\rceil\rceil\rceil$ is an element in the domain of $\inducN{t}{c} = \lceil c, \lceil d \rceil^3, \lceil \lceil e], \lceil f \rceil\rceil\rceil$.). All the elements $d, e$, and $f$ are leaves, as there is no element whose parent is either $d, e$, or $f$.  

The following definition shows how to extract groups from tree-like multisets. Groups are extracted via groups of multiset ingredients. 

\begin{defin}[{\bf Groups Associated with Tree-like Multisets}] \label{def:groups-treemult}
Let $t$ be a tree-like multiset over a set $F$. A set $G\subset \dom{\pure{F}{t}}$ is called a {\em group} associated with $t$ if there exists a group tree-like multiset $g\in \multing{t}$ such that $G = \{\rootm{x}: x\in \dom{g}\}$.  We define $\prntmult{t}{G} \eqdef \prntmult{t}{e}$ for an element $e\in G$ and call it the parent of $G$.\footnote{Note that $\forall e, e'\in G: \prntmult{t}{e} = \prntmult{t}{e'}$.}

The set of all groups associated with $t$ is denoted by $\agrp_t$. Let $\agrp(f)$ denote the set of all groups $G$ whose parent is $f$, \ie,  $\agrp(f) = \{ G\in \agrp_t: \prntmult{t}{G} = f \}$.
%
\qed 
\end{defin}

For example, consider the multiset $t = \lceil a, \lceil [b\rceil, \lceil c, \lceil d\rceil^4\rceil \rceil^5, \lceil e, \lceil f \rceil^3, \lceil \lceil g \rceil, \lceil h \rceil \rceil \rceil^2 \rceil$.  There are two  groups of tree-like multisets $g_1 = \lceil \lceil b\rceil, \lceil c, \lceil d\rceil^4\rceil \rceil$, ~$g_2 = \lceil \lceil g \rceil, \lceil h \rceil \rceil$. According to \defref{def:groups-treemult}, the groups corresponding to $g_1$ and $g_2$ would be, respectively, equal to the sets ~$G_1 = \{~\rootm{\lceil b\rceil}, ~\rootm{\lceil c, \lceil d\rceil^4\rceil}~\} = \{b, c\}$ and ~$G_2 = \{~\rootm{\lceil g \rceil}, ~\rootm{\lceil h \rceil} ~\} = \{g, h\}$.

We have already shown how to extract the corresponding tree and groups from a given tree-like multiset. All we need to do now is to know how to extract  multiplicities from tree-like multisets. The following definition shows how to do so.  
\begin{defin}[{\bf Multiplicities Associated with Tree-like Multisets}] \label{def:card-treemult}
For a given tree-like multiset $t\in \treemulthier(F)$ over a set $F$, we define a function $\acrd_t: \big( \dom{\pure{F}{t}} \setminus \{\rootm{t}\} \big) \cup \agrp_t \rarr \nat$ as follows: 
\begin{equation*}
\acrd_t(e) = 
\begin{cases}
|e| & \text{if $e\in \agrp_t$}\\
 \#_t(\inducN{t}{e})  & \text{otherwise}
\end{cases}  
\end{equation*} 
Recall that $\inducN{t}{e}$ and $\#_t(\inducN{t}{e})$ denote the tree-like multiset induced by $e$ and the multiplicity of $\inducN{t}{e}$ (see \defref{def:mult-ing}), respectively. 
%
\qed
\end{defin}
As an example, consider again the tree-like multiset  $t = \lceil a, \lceil b ]^5, \lceil c, \lceil d \rceil^3, \lceil \lceil e \rceil, \lceil f \rceil \rceil \rceil^2 ]$. It has only one associated group  $G =  \{ e, f \}$. According to \defref{def:card-treemult}, $\acrd_t$ is defined on $\{a, b, c, d, e, f, G\}$ as follows: 

$\acrd_t(b) =  \#_t(\inducN{t}{b})  =  \#_t( \lceil b \rceil )  = 5$.

$\acrd_t(c) =  \#_t(\inducN{t}{c})   = \#_t( \lceil c, \lceil d \rceil^3, \lceil \lceil e \rceil, \lceil f \rceil \rceil \rceil )  =  2$.

$\acrd_t(d) =   \#_t(\inducN{t}{d})\}  = \#_t(  \lceil d \rceil )  = 3$.

$\acrd_t(e) =  \#_t(\inducN{t}{e}) \}  =  \#_t(  \lceil e \rceil ) =  1$.

$\acrd_t(f) =  \#_t(\inducN{t}{f}) \}  =  \#_t(  \lceil f \rceil  ) =  1$.

$\acrd_t(G) =  \#_t(\inducN{t}{G})  =  |G| =   2$.

Now we are at the point where we can prove that any tree-like multiset is a hierarchical product of some CFD. 
\begin{theo}\label{th:treemult-2-CFD}
For any tree-like multiset $t$, there is a CFD $\ancd$ such that $t\in \products(\ancd)$ \qed
\end{theo}
 \begin{figure}[h]
\centering
    \includegraphics[scale=0.45]{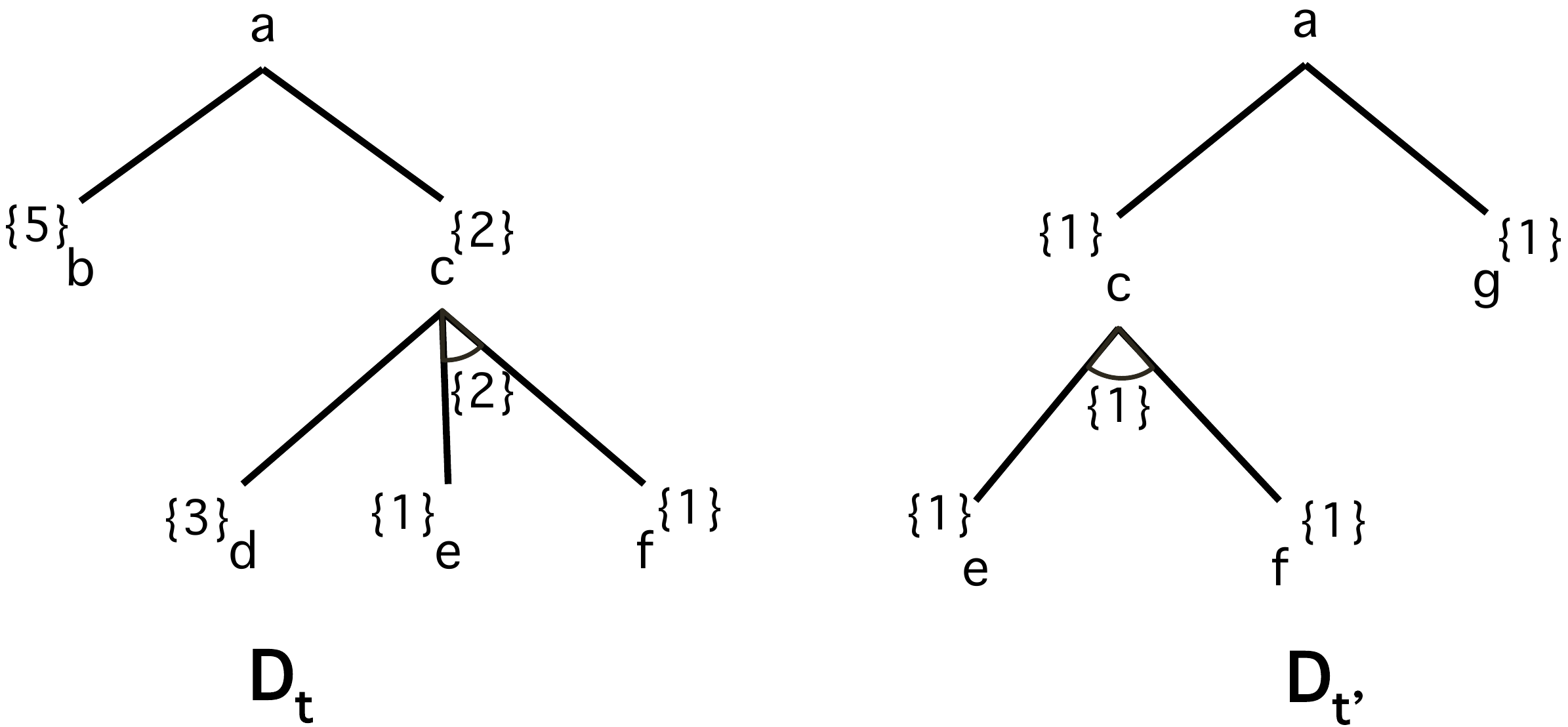}
\caption{Representative CFDs of single tree-like multisets: example}
\label{fig:treemult-to-CFD}
\end{figure}

For an example, consider  the tree-like multisets  $ t = \lceil a, \lceil b\rceil^5, \lceil c, \lceil d\rceil^3, \lceil\lceil e\rceil, \lceil f\rceil\rceil\rceil^2\rceil$ and $t' = \lceil a, \lceil c, \lceil \lceil e \rceil \rceil \rceil, \lceil g \rceil^3 \rceil$. The CFDs $\ancd_t$ and $\ancd_{t'}$ in \figref{fig:treemult-to-CFD} represent two CFDs whose hierarchical semantics include $t$ and $t'$, respectively. 


\section{Characterization of Hierarchical Semantics} \label{sec:merge-tree-mult}
In the previous section, we showed that a multiset is a hierarchical product of some CFDs if and only if it is a tree-like multiset.  In this section,  we want to see what sets of tree-like multisets can be the hierarchical theory of a CFD.  We first define the notions {\em mergeable tree-like} and {\em completely mergeable tree-like multisets}. A set of tree-like multisets is mergeable if it represents a subset of the hierarchical theory of some CFDs. It is called completely mergeable if it is equal to the hierarchical theory of a CFD.  

\begin{defin}[{\bf Mergeable Tree-like Multisets}] \label{def:mergeable-treemult}
We say that the elements of a (possibly infinite) set of tree-like multisets $U$ are 

(i) {\em mergeable} if there exists a CFD $\ancd$ such that $U \subseteq \products(\ancd)$. We then call $\ancd$ a {\em representative} CFD of  $U$. 

(ii) {\em completely mergeable} if there is a CFD $\ancd$ such that $U = \products(\ancd)$. 
 \qed
\end{defin}

\begin{figure}[h]
\centering
    \includegraphics[scale=0.65]
                 {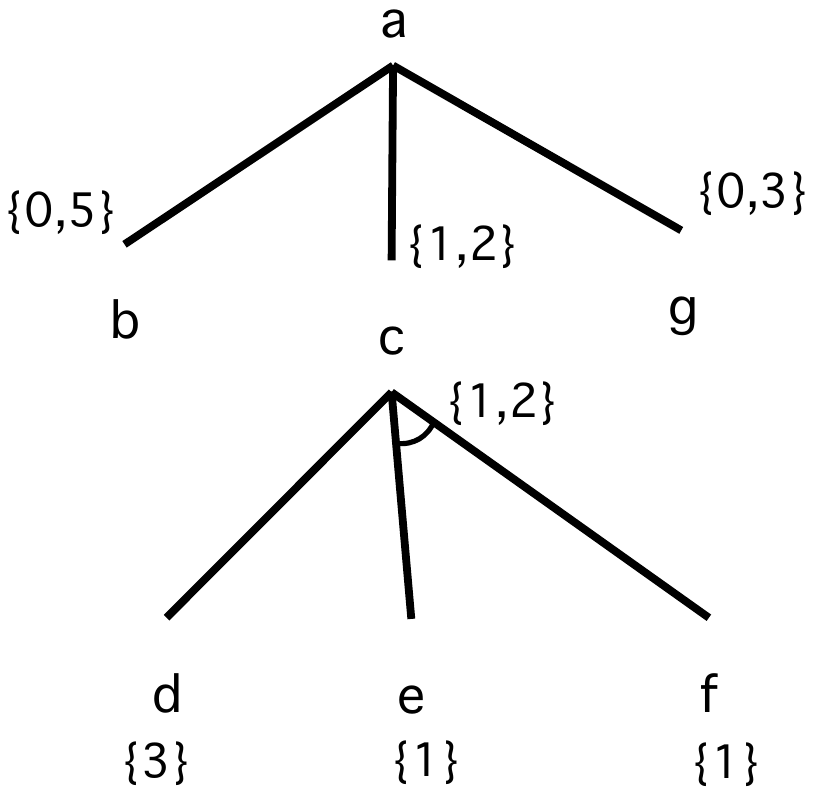}
\caption{Representative CFDs of mergeable tree-like multisets: example}
\label{fig:merged-treemult}
\end{figure}

According to \theoref{th:treemult-2-CFD}, any singleton set of tree-like multisets is mergeable.  Consider the tree-like multisets  $t = \lceil a, \lceil b\rceil^5, \lceil c, \lceil d\rceil^3, \lceil \lceil e\rceil, \lceil f\rceil\rceil\rceil^2\rceil$ and $t' = \lceil a, \lceil c, \lceil \lceil e \rceil \rceil \rceil$ $, \lceil g \rceil^3 \rceil$. \figref{fig:merged-treemult} represents a CFD whose hierarchical theory includes $t$ and $t'$.  Therefore, they are mergeable. However, $t$ and $t'$ are not completely mergeable. 

As a simple example of  non-mergeable tree-like multisets, consider $n = \lceil a, \lceil b\rceil^3 \rceil$ and $n' = \lceil b, \lceil a\rceil^2 \rceil$. They are not mergeable, as their roots  are different.

There is no unique CFD representing a given set of tree-like multisets. For example, replacing the multiplicity domain of node $b$ in \ancd\ (\figref{fig:merged-treemult}) by any other multiplicity domains including $0$ and $5$ (\eg, $\nat$), the CFD would still represent $t$ and $t'$. Another example:  adding an optional subfeature\footnote{multiplicity domain with lower bound 0} to the node $b$, the CFD is still a representative of $t$ and $t'$. Indeed, for a given set of mergeable tree-like multisets, there is an infinite number of representative CFDs. Therefore, a notion of {\em minimality} for representative CFDs can be useful.  

\begin{defin}[{\bf Minimal Representative CFDs}] \label{def:smallest-cfd}
A CFD \ancd\ is called a {\em minimal representative} CFD of a given set of mergeable tree-like multisets $U$ if

(i) it is a representative CFD of $U$, and 

(ii) for any other representative CFD $\ancd'$ of $U$, $|\products(\ancd)| \leq |\products(\ancd')|$. 
\\
Let $\mergcfds{U}$ denote the family  of minimal representative CFDs of $U$.
%
 \qed
\end{defin}

\begin{figure}[h]
\centering 
    \includegraphics[scale=0.65] 
                 {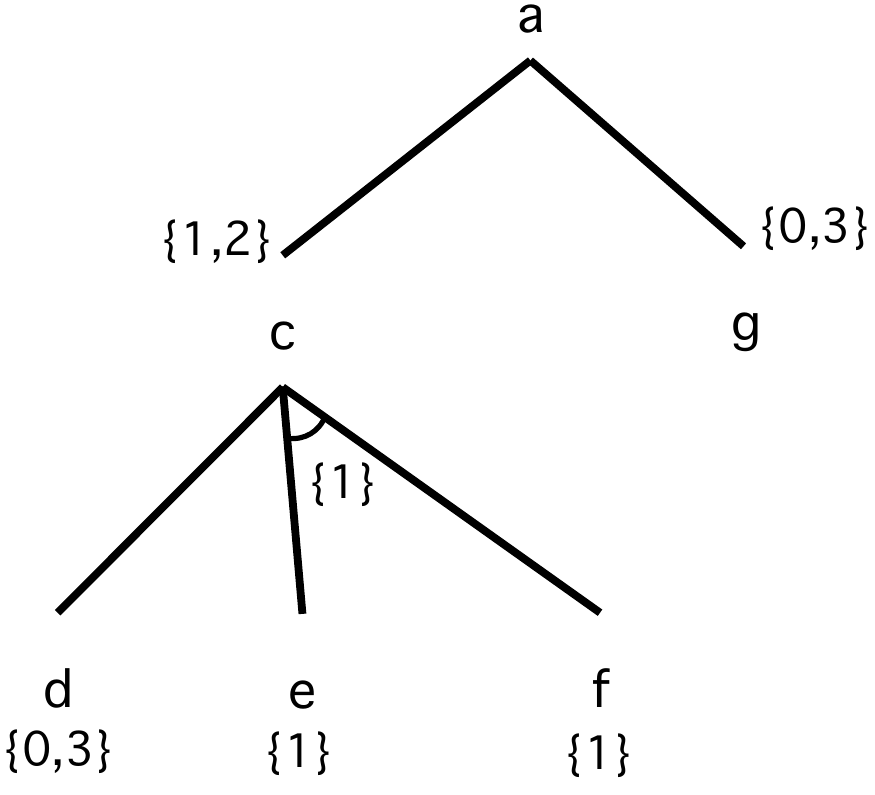}
\caption{Representative CFDs of mergeable tree-like multisets: example} 
\label{fig:merged-treemult2}
\end{figure}

The CFD $\ancd$  in \figref{fig:merged-treemult} represents a minimal representative CFD of the tree-like multisets $t = \lceil a, \lceil b\rceil^5, \lceil c, \lceil d\rceil^3, \lceil\lceil e\rceil, \lceil f\rceil\rceil\rceil^2\rceil$ and $t' = \lceil a, \lceil c, \lceil \lceil e \rceil \rceil \rceil, \lceil g \rceil^3 \rceil$. For these two tree-like multisets, there is, indeed, only one minimal representative CFD. Now, consider another tree-like multiset $t'' = \lceil a, \lceil c, \lceil d\rceil^3, \lceil\lceil e\rceil\rceil\rceil^2 \rceil$. A minimal representative CFD of $t''$ and $t'$ is represented in  \figref{fig:merged-treemult2}. However, this is not the only minimal CFD representing these two tree-like multisets: replacing $f$ by another feature, say $x$, we obtain another minimal representative CFD of $t''$ and $t'$.  

%
\begin{remar}
Note that, according to  \theoref{th:unique}, there is a single minimal representative CFD of  a given set of completely mergeable tree-like mutlisets.   \qed
\end{remar}

In the rest of this section, we are going to characterize  mergeable tree-like multisets. 
To this end, we first introduce the notion of mergeable trees. 
\begin{remar}
Note that a mergeable set of tree-like multisets may be infinite. However, it is always enumerable, as the hierarchical theory of a CFD is always enumerable. This simple fact is used in the following definitions and theorems. 
\qed   
\end{remar}

\begin{defin}[{\bf Mergeable Trees}] \label{def:merge-tree-treemult}
Consider an enumerable set of trees $\treefam =  \{T_i: i\in I\}$, where $I$ enumerates its elements. Let $T_i = (N_i, r_i, {\_}^{\uparrow_i})$, $\forall i \in I$. We say that the trees in $\treefam$ are mergeable if 

(i) $\forall  i, j \in I: r_i = r_j$.

(ii) $\forall  i, j \in I, \forall n\in (N_i\cap N_j)\setminus \{r_1\} : n^{\uparrow_i} = n^{\uparrow_j}$. 
\\
Then the tuple $(N, r, \prnt{\_})$, where $ N = \bigcup_{i \in I} N_i, r=r_1\text{, and }  \prnt{\_} = \bigcup_{i \in I} {\_}^{\uparrow_i}$ is a tree. We use the notation $\merg{\treefam}$ to denote this tree and call it the {\em representative tree} of $\treefam$. 
%
\qed
\end{defin}

\begin{figure}[h]
\centering
    \includegraphics[scale=0.6]{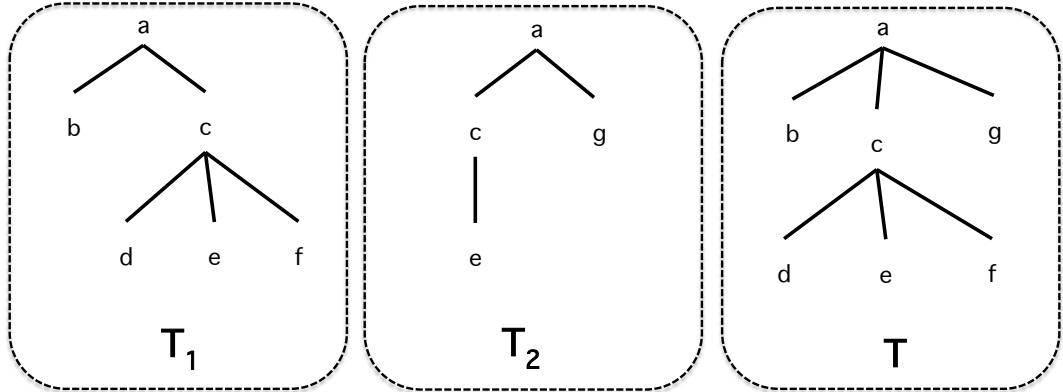}
\caption{Megeable trees and their representative trees: an example}
\label{fig:merge-trees}
\end{figure}

 As an example,  consider the megeable trees $T_1, T_2$ and their representative tree $T$ in \figref{fig:merge-trees}.    
Taking advantage of this notion, we characterize  mergeable tree-like multisets in the following theorem. 

\begin{theo} \label{th:merg-2treemults}
Consider an enumerable set of tree-like multisets $U = \{t_i: i\in I\} \subset \treemulthier(F)$ over a set $F$, where $I$ enumerates its elements. 
  Let $T_i = (N_i, r_i, {\_}^{\uparrow_i})$ and $\agrp_i$  ($\forall i\in I$) denote the $t_i$'s associated tree and groups, respectively (see Definitions \ref{def:tree-treemult} and \ref{def:groups-treemult}, respectively).  
%
%
 The tree-like multisets in $U$ are mergeable iff:

(i) $\forall i, j \in I: T_i, T_j$ are mergeable.

(ii) $\forall i,j\in I, \forall n\in N_i\cap N_j: (\exists G\in \agrp_i: n\in G) \implies (\exists G\in \agrp_j: n\in G)$.
\qed
\end{theo}

The above theorem characterized mergeable tree-like multisets. 
However, it does not lead us to a  pragmatic approach when a given set of tree-like multisets is infinite. 
We need to address this problem.  
Note that what makes the hierarchical theory of a CFD infinite is due to some infinite multiplicity domains of some nodes, \eg,  the multiplicity domain $\nat \setminus \{0,1,6\}$ on $\axle$ in the CFD in \figref{fig:CFD-intro}. However, as we saw in \theoref{th:merg-2treemults}, multiplicities on elements in tree-like multisets have no influence in making them mergeable or not.  
 We will use this clue to address the problem.  We first introduce the notion of {\em relaxed multisets}. A relaxed version of a given multiset is obtained by changing all multiplicities of its ingredients to $1$. For an example, the relaxed multiset of $\lceil a, \lceil b\rceil^5, \lceil c, \lceil d\rceil^3, \lceil\lceil e\rceil, \lceil f\rceil\rceil\rceil^2\rceil$ would be $\lceil a, \lceil b\rceil, \lceil c, \lceil d\rceil, \lceil\lceil e\rceil, \lceil f \rceil\rceil\rceil\rceil$.

\begin{defin}[{\bf Relaxed Multisets}] \label{def:relax-mult}
Given a multiset $m\in \multhier(A)$ over a set $A$, its {\em relaxed multiset}, denoted by $\relaxm{m}$, is defined as follows: 

$
~~~~~\dom{\pure{A}{\relaxm{m}}} = \dom{\pure{A}{m}},
$

$
~~~~~\multing{\relaxm{m}} = \multing{m},
$

$
~~~~~\forall e\in \dom{\relaxm{m}}: \relaxm{m}(e) = 1,
$
 
$
~~~~~\forall n\in \multing{\relaxm{m}}, \forall e\in \dom{n}: n(e) = 1.
$
\\
For a given set of  multisets $U$, let $\relaxm{U}$ denote the set $\{\relaxm{m}: m\in U\}$. 
%
%
\qed
\end{defin}
The following proposition follows easily. 
\begin{propo} \label{prop:relax-cfd}
Let $T_t = (N_t, r_t, {\_}^{\uparrow_t}), \agrp_t, \acrd_t$ denote tree, groups, and multiplicities associated with a given tree-like multiset $t \in \treemulthier(F)$ for a set $F$  (see Definitions \ref{def:tree-treemult}, \ref{def:groups-treemult}, and \ref{def:card-treemult}.). The tree and groups associated with $\relaxm{t}$ are equal to  $T_{\relaxm{t}} = T_t$ and $\agrp_{\relaxm{t}} = \agrp_t$, respectively. The multiplicities associated with $\relaxm{t}$, \ie, $\acrd_{\relaxm{t}}$, is defined as follows:
%
\begin{equation*}
\acrd_{\relaxm{t}}(e) = 
\begin{cases}
\{1\} & \text{if } e\in (N_t \setminus \{r_t\})\\
\acrd_t(e) & \text{if } G\in \agrp
\end{cases}  
\end{equation*}
\qed
\end{propo}  

%
To specify whether a given set of tree-like multisets is mergeable or not, we just need to deal with its relaxed version (see \theoref{th:merge-relax-yn}(i)).  More interestingly (and practically useful), the relaxed version of a set of mergeable tree-like multisets is finite (see \theoref{th:merge-relax-yn}(ii)). 
\begin{theo}  \label{th:merge-relax-yn}
Consider an emeumerable set of tree-like multisets $U \subset \treemulthier(A)$ over a set $A$.

(i) $U$ is mergeable iff $\relaxm{U}$ is. 

(ii) $U$ is mergeable implies that $\relaxm{U}$ is finite. 
\qed
\end{theo}

Now, we want to characterize completely mergeable tree-like multisets. This is done in \theoref{th:charc-complete-merge}. Before getting to the theorem, let us see some examples. Consider the set $U = \{t_1, t_2, t_3, t_4\}$ of tree-like multisets, where $t_{1-4}$ are as follows:

~~~~~ $t_1 = \lceil a, \lceil b\rceil^5, \lceil \lceil c \rceil \rceil \rceil$, 

~~~~~ $t_2 = \lceil a, \lceil b\rceil^5, \lceil \lceil d \rceil^3 \rceil \rceil$, 

~~~~~ $t_3 = \lceil a, \lceil b\rceil^2, \lceil\lceil c\rceil\rceil \rceil$, 

~~~~~ $t_4= \lceil a, \lceil b \rceil^2, \lceil \lceil d\rceil^3 \rceil \rceil$.  
\\
Their relaxed multisets are represented in the following (as usual, we denote the set of the following tree-like multisets by $\relaxm{U}$):

~~~~~ $\relaxm{t_1} = \relaxm{t_3} = \lceil a, \lceil b\rceil, \lceil \lceil c\rceil \rceil \rceil$, 

~~~~~ $\relaxm{t_2} = \relaxm{t_4} = \lceil a, \lceil b \rceil, \lceil \lceil d \rceil \rceil \rceil$.
%
%
\\

\begin{figure}[h]
\centering
       \includegraphics[scale=0.5]
                 {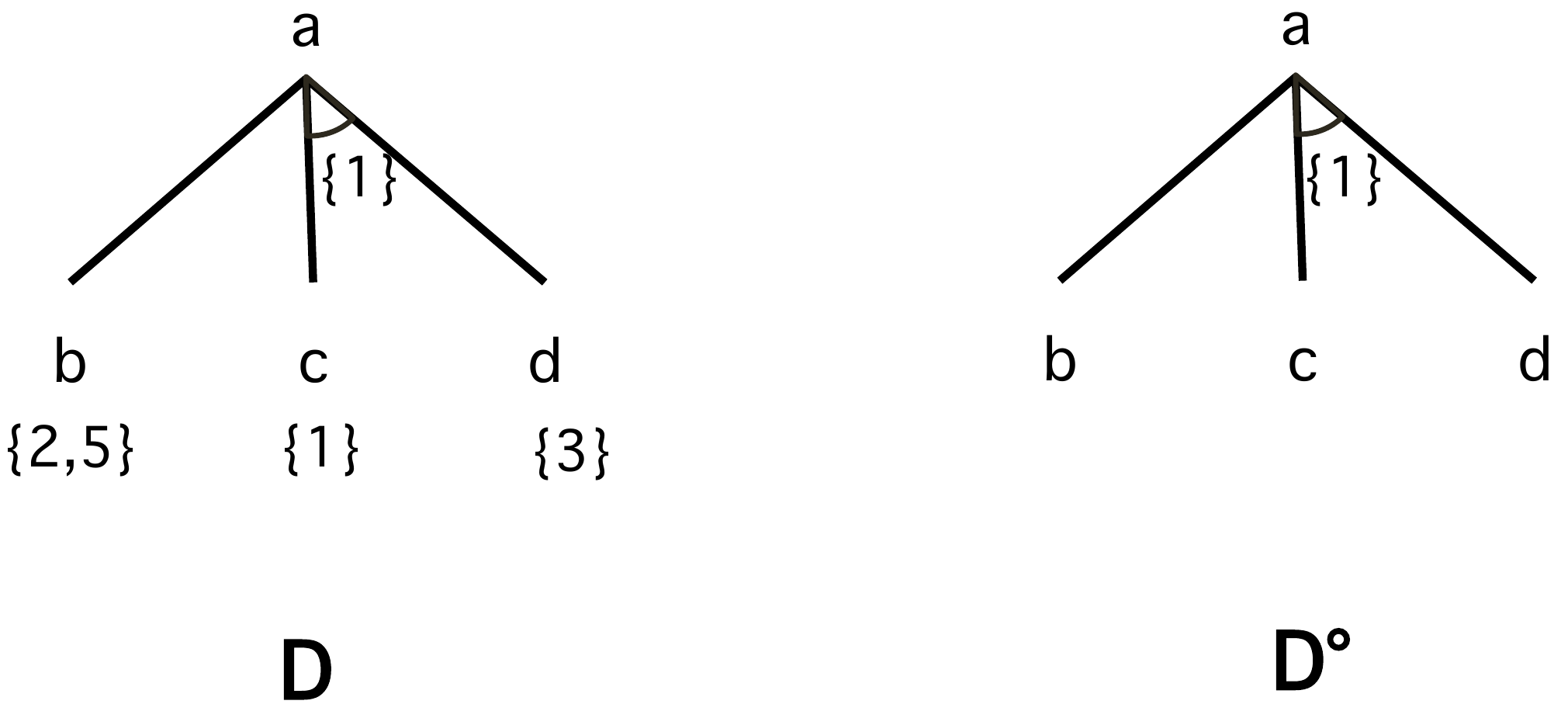}
\caption{Minimal representative CFDs of  $U$ and $U^\circ$} 
\label{fig:complete-merge} 
\end{figure}

Two minimal representative CFDs of  $U$ and $\relaxm{U}$ are represented in \figref{fig:complete-merge} as \ancd\ and $\relaxm{\ancd}$, respectively.\footnote{Since $U$ is completely mergeable, there is only one minimal representative CFD of $U$.}  Since $\products(\ancd) = U$ and $\products(\relaxm{\ancd}) = \relaxm{U}$, both $U$ and $\relaxm{U}$ are completely mergeable.  

\begin{figure}[h]
\centering
       \includegraphics[scale=0.5]
                 {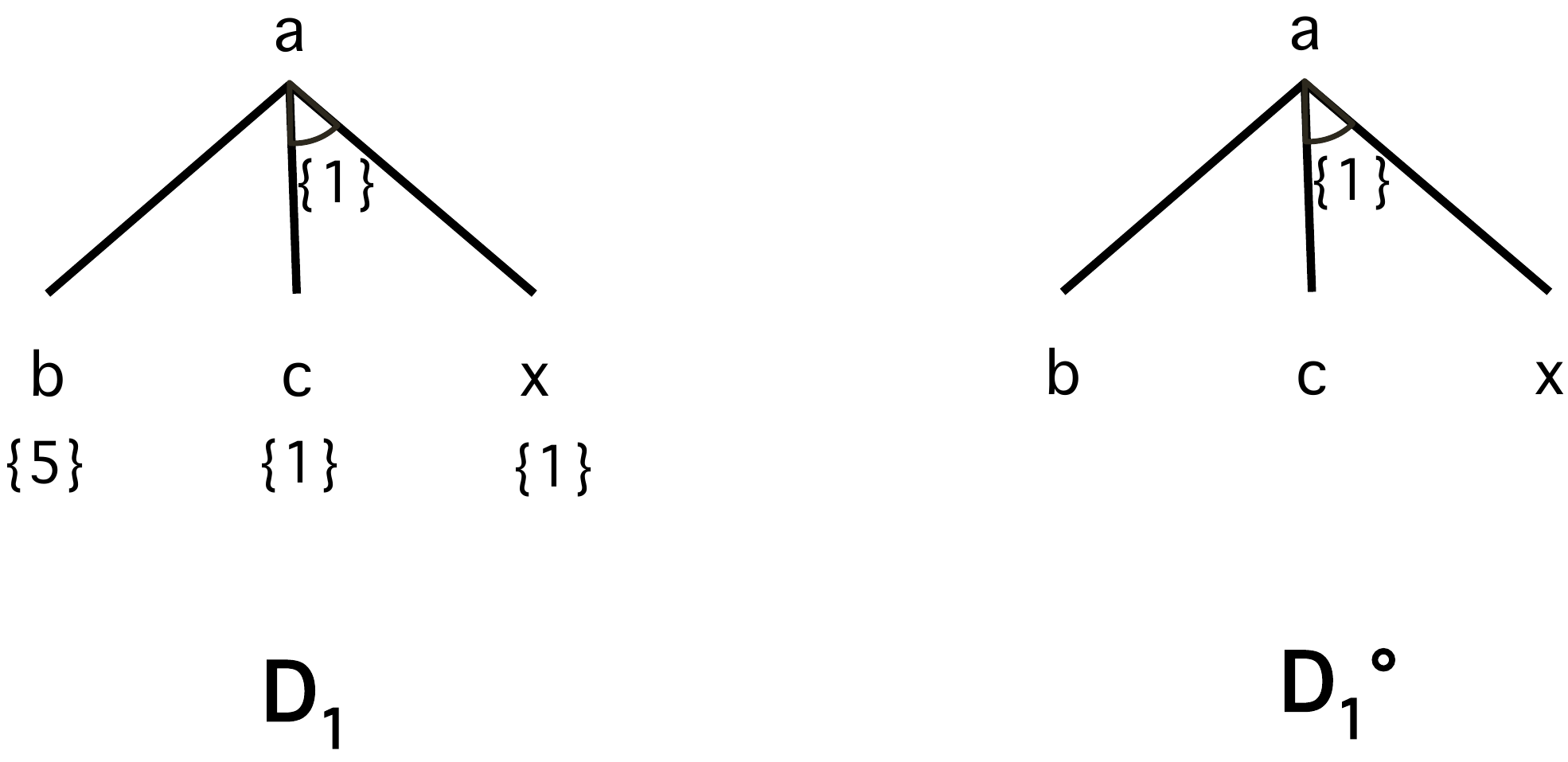}
\caption{Minimal representative CFDs of  $U_1$ and ${U_1}^\circ$} 
\label{fig:complete-merge-not} 
\end{figure}

Now, consider $U_1 = U \setminus \{t_2, t_4\}$. Clearly, $U_1$ is not completely mergeable. A minimal representative CFD $\ancd_1$ of $U_1$ is represented in \figref{fig:complete-merge-not}, where $x$ can be any feature not equal to $a, b$, or $c$. A representative CFD $\relaxm{\ancd_1}$ of $\relaxm{U_1}$ is also represented in \figref{fig:complete-merge-not}. Note that $\relaxm{U_1}$ is not completely mergeable either. 

What we saw in the above examples is indeed a general rule: For any completely mergeable tree-like multisets $U$, $\relaxm{U}$ would be completely mergeable too. However, we cannot characterize completely mergeable multisets relying on just their relaxed multisets. Indeed, there are  sets of tree-like multisets which are not completely mergeable, but their relaxed multisets are. As an example, consider the set of multisets $U_2 = U \setminus \{t_3\}$. Clearly, $U_2$ is not completely mergeable. The CFD \ancd\ in \figref{fig:complete-merge} is a minimal representative CFD of $U_2$ (recall that it is also a representative CFD of $U$). Since $\relaxm{U_2} = \relaxm{U}$,  $\relaxm{U_2}$ is completely mergeable (as $\relaxm{U}$ is).  

The above discussion shows that characterization of completely mergeable tree-like multisets goes via their relaxed tree-like multisets and multiplicities. We will need the following notion. 
\begin{defin} [{\bf Overall Multiplicities}] \label{def:overallmult}
Given a set of tree-like multisets $U \subset \treemulthier(F)$ for a set $F$, we define a function $\acrd_U : F \rarr 2^\nat$ as follows: 
$
\acrd_U(f) = \bigcup_{t\in U} \{ \#_t(\inducN{t}{f}) \}
$.\footnote{Recall that $\inducN{t}{f}$ and $\#_t(\inducN{t}{f})$ denote the multiset induced by $a$ and the multiplicity of $\inducN{t}{f}$ in $t$, resp.}
\qed
\end{defin}
\begin{theo} \label{th:charc-complete-merge}
Consider an enumerable set of tree-like multisets $U \subset \treemulthier(F)$ over a set $F$. $U$ is completely mergeable iff

(i) $\relaxm{U}$ is completely mergeable, and

(ii) $\forall t\in \relaxm{U}, \forall f\in \dom{\pure{F}{t}}, \forall c\in \acrd_{U}(f), \exists t'\in U: (\relaxm{t'} = t) \wedge (\#_{t'}(\inducN{t'}{f}) = c)$. \qed
\end{theo}

In \theoref{th:unique}, we showed that two CFDs are equal iff their hierarchical theories are equal. This implies that there is a unique minimal representative CFD of given completely mergeable tree-like multisets. Thus, we get to the following statement, which is a corollary of Theorems \ref{th:charc-complete-merge} and \ref{th:unique}. 
\begin{corol} \label{corol:bijection}
There is a bijection between the domains of CFDs and completely mergeable tree-like multisets. \qed
\end{corol}


\section{Other Practical Applications}\label{sec:practice}
 The hierarchical theories of CFDs could  also be used in the {\em reverse engineering} of CFDs (an important problem in feature modeling), as the hierarchical theory of a given CFD captures all information about the CFD. In \secref{sec:tree-mult} and \secref{sec:merge-tree-mult}, we characterized the hierarchical products and semantics of a given CFD, respectively. The proofs given for corresponding theorems are all constructive: \theoref{th:treemult-2-CFD} constructively shows that there is a CFD representing a given tree-like multiset; \theoref{th:merge-relax-yn}, whose proof is constructive, characterizes mergeable tree-like multisets;  \theoref{th:charc-complete-merge}, whose proof constructively shows
how to retrieve the CFD from its hierarchical semantics, characterizes completely mergeable multisets.

Another important application of the hierarchical semantics of CFDs regards {\em feature model management}, which is  an active area in feature modeling. By feature model management, we mean feature model composition via some operations like merging, intersection, and union, etc \cite{segura2008, acher2010composing, acher2013}. Characterization of the hierarchical semantics come in handy here. As an example, suppose that we want to obtain the merge  of two CFDs $\ancd_1$ and $\ancd_2$. We need to address the two following questions: Are $\ancd_1$ and $\ancd_2$ mergeable? What would be the result of their merge, if they are mergeable? To address these questions, we  first obtain their hierarchical semantics $\products(\ancd_1)$ and $\products(\ancd_2)$, respectively.  We then decide whether their union is mergeable or not. To this end, we take advantage of \theoref{th:merge-relax-yn}. This would address the first question. If they are mergeable, then we obtain a representative CFD of $\products(\ancd_1) \cup \products(\ancd_2)$. The proof of \theoref{th:merge-relax-yn} constructively shows how to obtain a representative CFD of a set of mergeable  tree-like multisets. As for the intersection (union, respectively) of $\ancd_1$ and $\ancd_2$, we first obtain the intersection (union, respectively) of their hierarchical semantics and then decide whether the obtained set of tree-like multisets is completely mergeable or not. To this end, we would apply \theoref{th:charc-complete-merge}.  


\section{Related  Work} \label{sec:related}
\subsection{Flat Semantics}
The most well-known formulation of flat semantics of CFDs was given via context-free grammars by Czarnecki \etal\ in  \cite{czarnecki2005}: a given CFD is transformed to a context-free grammar and then the {\em multiset interpretation} of the corresponding language is considered as the semantics of the CFD. Formally, the authors mean the {\em Parikh image} \cite{parikh1961} of the language by its multiset interpretation: The Parikh image (a.k.a. Parikh vector) of a given word $w\in \Sigma^*$ ($\Sigma = \{\sigma_1, \ldots, \sigma_n\}$ denotes an alphabet) is the vector $(o_1, \ldots, o_n)$ where $o_i$ denotes the number of occurrences of $\sigma_i$ in $w$. Clearly, the Parikh image of a word can be expressed as a multiset over the alphabet. According to the informal description of what the authors provided in \cite{czarnecki2005}, this semantics must be equal to the flat semantics of the CFD, though it was not proven formally. In addition, the syntax defined for CFDs in \cite{czarnecki2005} has two restrictions on multiplicity domains of groups: (i) the multiplicity domain of a grouped feature is always $\{1\}$ and (ii) the multiplicity domain on a  group is a singleton pair of natural numbers. This restrictions have been relaxed in our formal framework without essentially complicating it.  

Another set theoretic definition of flat semantics can be found in \cite{safilian2015}. The syntax of CFDs formalized in \cite{safilian2015} supports labelled CFDs, \ie, CFDs in which the labels  of several nodes can be the same. 
In the present paper, we have considered unlabelled CFDs. However, our work can be easily transformed to labelled CFDs, using labelling functions on  flat and hierarchical multiset theories. 

Several formal semantics have been proposed for capturing the flat semantics of basic feature modeling, including a propositional logic encoding  \cite{mannion2002}, first order logic encoding \cite{sun2005},  algebraic  based semantics \cite{hofner2011}, context-free grammar encoding  \cite{deJonge2002}, generic semantics \cite{schobbens2007}.  

\subsection{Hierarchical Semantics}
The closest work to ours is \cite{safilian2015}, where a {\em faithful} semantics for CFDs was provided by using regular languages as the semantic domain. By a faithful semantics for a CFD, the authors mean a semantics capturing the flat semantics and the hierarchy of the CFD (all information essential for answering the existing analysis questions about the CFD). They first proposed a generalization of CFDs, called {\em cardinality-based regular expression diagrams} (CRDs) in which a label of a node can be any regular expressions built over a set of features. Then, a reduction process was provided  going from a given CRD to a regular expression. The authors proved that the regular expression generated for a given CFD provides a faithful semantics for the CFD. 
The similarity between our hierarchical multiset theory and this language based semantics of a given CFD is that both provides a faithful semantics for the CFD. However, there is a subtle difference between the faithfulness of the hierarchical theory and the language semantics. The hierarchical mutliset theory explicitly distinguishes between grouped and solitary features, while the language semantics does not. 
\begin{figure}[h]
\centering
    \includegraphics[scale=0.7]
                 {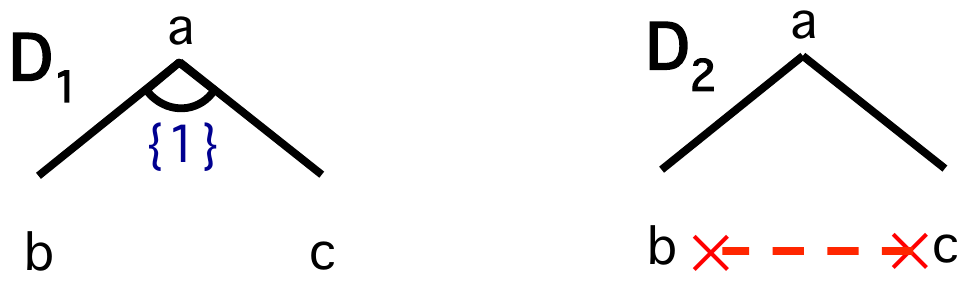}
\caption{Faithfulness in multiset and language semantics}
\label{fig:faith-related}
\end{figure}

As a simple example, consider the CFDs \ancd$_1$ and \ancd$_2$ in \figref{fig:faith-related}. The $\times$-ended arc between $b$ and $c$ in $\ancd_2$ denotes an exclusive constraint between them. According to \cite{safilian2015}, the language semantics of $\ancd_1$ and $\ancd_2$ would be the same $\{ab, ac\}$. However, their hierarchical theories would be, respectively, $\{\lceil a, \lceil \lceil b \rceil \rceil \rceil, \lceil a, \lceil \lceil c \rceil \rceil \rceil\}$ and $\{\lceil a, \lceil  b  \rceil \rceil, \lceil a, \lceil  c \rceil \rceil\}$. Therefore, the language semantics does not capture the differences between $\ancd_1$ and $\ancd_2$, while the hierarchical multiset semantics does. This shows that, although the language semantics of a CFD adequately captures the essential information for what we need to address the current practical analysis questions about CFDs, it looses some other information which may be in handy for other purposes, \eg, for reverse engineering of CFDs. 

\begin{figure}[h]
\centering
\includegraphics[scale=0.8]{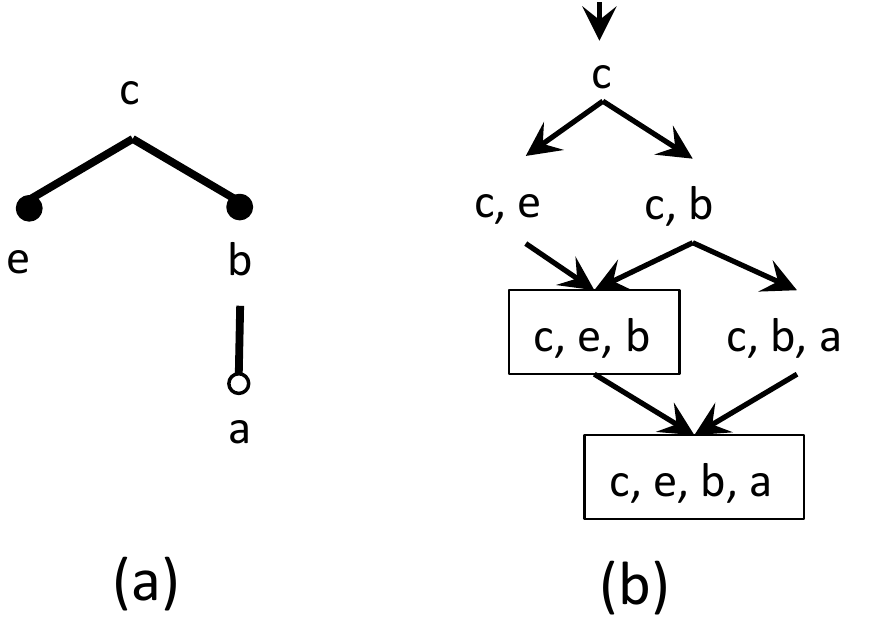}
\caption{(a) an FM \bfanfm, (b) PPL(\bfanfm)  
\label{fig:PPL-relation}}
\end{figure}

Diskin \etal\ in \cite{diskin2015} proposed a relational semantics for basic feature modeling. The structure  corresponding to a given BFD is called the {\em Partial Product Line} (PPL) of the BFD. The states of this structure are called {\em partial products}, which are sets of features satisfying the {\em exclusive constraints} (a partial product must not violate the exclusive constraints), {\em subfeature relationship} (a feature cannot be included in a partial product if its parent feature is not), and the {\em instantiation-to-completion} (\itoc) principle (processing a new branch of the feature tree should only begin after processing of the current branch has been completed). The initial state is a singleton set $\{r\}$ where $r$ is the root feature. The flat semantics (called product line in \cite{diskin2015}) of the BFD is a subset of the set of states (partial products).  \figref{fig:PPL-relation}(a) is a BFD and its PPL is represented in \figref{fig:PPL-relation}(b). In this figure, the states representing the flat products of the BFD are boxed. 
  {\em Singletonicity} is one of the important properties of PPLs. This property says that if there is a transition $P\lra P'$ between two products $P$ and $P'$, then $P' = P \cup \{f\}$ for some feature $f\not\in P$. 
 
The authors also propose a CTL based logic for specifying PPLs and show how to transform a given BFD to a complete logical theory of its PPL. One similarity between the PPL and the hierarchical  theory of a given BFD is that both captures the flat semantics and the hierarchy of the BFD (the hierarchical theory of the BFD in \figref{fig:PPL-relation}(a) is the set $\{ \lceil c, \lceil e \rceil, \lceil b \rceil \rceil, \lceil c, \lceil e \rceil, \lceil b, \lceil a \rceil \rceil \rceil \}$). However, like we discussed above for the language semantics, the hierarchical theory explicitly distinguishes between the grouped and solitary features, while the PPL does not.  

A process calculi, called {\em PL-CCS},  developed in  \cite{leucker2012, gruler2008}  for modeling the behaviour of PLs. PL-CCS extends the classical CCS by an operator $\oplus$  to model variability.  $\oplus$ is a kind of choice applied at well-defined variation points. Each $\oplus$ occurence in a PL-CCS expression is equipped with a unique index, and runtime occurrences with the same index must make the same choice. 
This differentiates $\oplus$'s behaviour  from the classical non-deterministic choice  in CCS.  In PL-CCS, processes are interpreted as products.  The behaviour of a product line is given by a set of process definitions whose semantics is given by multi-valued Kripke structures. Importantly, PL-CCS allows for recursive definitions of processes, which makes them even applicable for cardinality-based feature modeling. Three kinds of semantics are given via multivalued Kripke structures for a PL-CCS program: {\em flat}, {\em unfolded}, and {\em configured transitions}. The flat semantics is a set of transitions systems each of which models a full product. The unfolded semantics is a single transition system modeling the whole PL. In the configured transition semantics, the states are identified by configurations. We think that there might be a tight  relation between the hierarchical semantics and the set of PL-CCs' flat semantics. This research task has been left as a future work. 

\section{Conclusion} \label{sec:conclusion}
The flat theory is commonly considered as the semantics of CFDs in the literature.  
In this paper, we have provided two formal definitions for flat semantics including  a recursive one. Therefore, deciding whether a given multiset is a valid flat product for a given CFD or not is algorithmic.  The flat theory of a given CFD can address a large number of analysis questions about the CFD. However, it does not capture all useful information about the CFD. To overcome this problem, we have proposed another multisets-based semantics for CFDs, called the hierarchical semantics. 

To define the hierarchical theory of a given CFD, we first defined a hierarchy of multisets over the set of features whose first class is the set of finite multisets over the features and other classes are defined as the set of all finite multisets built over the union of the previous classes.  A hierarchical product of a CFD is defined as a multiset (in the corresponding multisets hierarchy) such that its rank  is given by  the depth of the CFD and the multiplicities satisfy the multiplicity constraints of the CFD. The set of all hierarchical products is called the hierarchical theory of the CFD. We have proven that the hierarchical theory of a CFD captures all information of the CFD so that one can get back to the CFD from its hierarchical semantics. This also means that one can address any question about the CFD based on its hierarchical semantics. It is easy to see that deciding whether a given multiset is a hierarchical product of a given CFD or not is algorithmic (see the recursive definition of hierarchical products in \defref{def:hier-products}).

We have proven that there is a bijection between flat and hierarchical semantics of a given CFD, \ie, a hierarchical product is a hierarchical version of a flat product. 

To characterize a multiset being a hierarchical product of a CFD, we proposed the notion of {\em tree-like multisets}: We have proven that a multiset can be a hierarchical product of some CFDs iff it is a tree-like multiset. Also, we have characterized a set of tree-like multisets being the hierarchical theory of a CFD.

We have proven that the hierarchical theory of a CFD provides the most faithful semantics. Indeed, one can get back to the CFD from its hierarchical theory.  We have also discussed several possible practical applications of the mutliset theories of CFDs. 


\bibliographystyle{abbrv}
\bibliography{refs}

\appendix 

\section{Proofs} \label{sec:app}
The following definition will be used in the proof of \lemref{th:flat-products}. 

\begin{defin}[{\bf Upper  Diagram Induced by Depth}]  \label{def:induce-depth}
Let $\ancd = (F, r, \prnt{\_}, \agrp, \acrd)$ be a CFD and $1 \leq k\leq \depth{\ancd}$. The {\em upper diagram induced  by $k$} is a CFD $\inducD{\ancd}{k} = (F', r, \prnt{\_}|_{F'}, \agrp',$ $\acrd')$,  where $F' = \{f\in F: \depth{n} \leq k\}$, $\agrp' = \agrp \cap 2^{F'}$, and $\acrd' = \acrd|_{F'\uplus \agrp'}$, \ie, its tree is a subtree of \ancd's tree where the nodes are in depth less than or equal to $k$; all other components are inherited from $\ancd$.
 %
\end{defin}

\begin{figure}[h]
\centering
    \includegraphics[scale=0.7]
                 {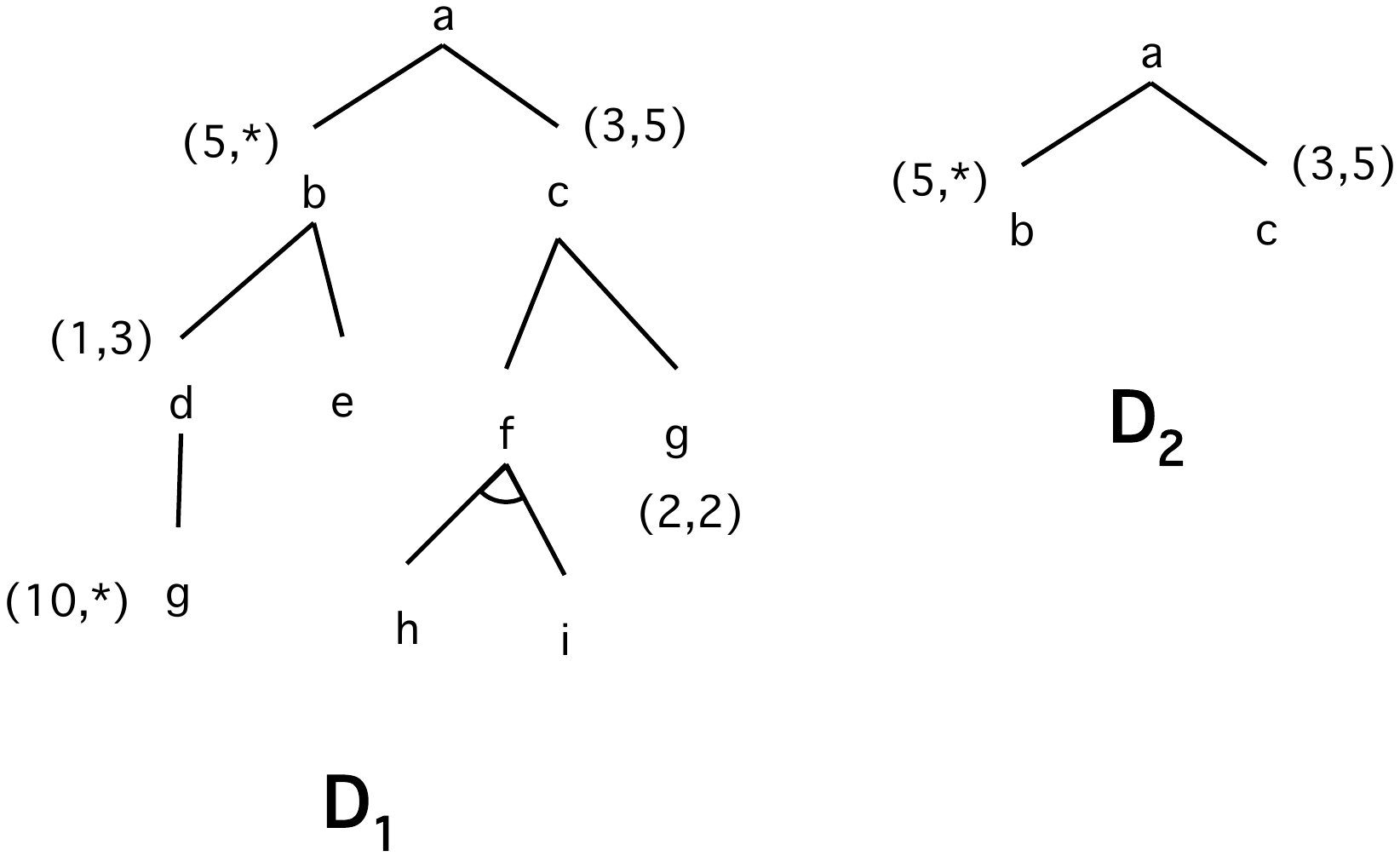}
\caption{\ancd$_2$: Upper diagram induced by depth 2 of $\ancd_1$}
\label{fig:induce-depth}
\end{figure}

For example,  $\ancd_2$ is the upper  diagram induced by depth 3 of $\ancd_1$ in \figref{fig:induce-depth}.
~\\
~\\
{\bf Lemma \ref{th:flat-products}.} 
Given a CFD $\ancd = (F, r, \_^\uparrow, \agrp, \acrd)$, for any multiset $m$ over $F$:  $m \in \pproducts(\ancd)$ iff $m$ satisfies the following conditions:

(i)  $m(r) = 1$,

(ii) $\forall f\in \asol \cap \chld{r},~ \exists c\in \acrd(f),~ \exists n\in \pproducts(\inducN{\ancd}{f}),~ \forall e\in \dom{n}: m(e) = c\times n(e)$.

(iii) $\forall G\in \agrp \cap 2^{\chld{r}},~ \exists n \in \pproducts(\ancd, G),~ \forall e\in \dom{n}: m(e) = n(e)$.
\begin{proof}
For any CFD $\ancd$ and any flat multisets $m$ over $F$, we show that both the following statements hold:
 
 (1) $m\in \pproducts(\ancd) \implies m$ satisfies Th-(i), (ii), and (iii).\footnote{Th-(i), (ii), and (iii) are abbreviations for \theoref{th:flat-products}(i), (ii), and (iii), respectively.} 
 
 (2) $m$ satisfies Th-(i), (ii), (iii) $\implies m\in \pproducts(\ancd)$. 
 
{\em \underline{Proof of (1)}}: \\
We prove (1) by the following inductive reasoning on the depth of CFDs.  

({\em base case}): Consider a CFD \ancd\ with  $\depth{\ancd} = 1$ and $r$ as its root, \ie, $F_\ancd= \{r\}$ and any other components are empty.  
The only flat product is $m =\lceil r\rceil$. Holding each of the conditions Th-(i), (ii), and (iii) follows obviously, as $m(r) = 1$, $\asol \cap \chld{r}=\varnothing$, and $\agrp \cap 2^{\chld{r}} = \varnothing$. 

({\em hypothesis}): Assume that for any CFD \ancd\ with $\depth{\ancd} < k$ (for some $k$), any $m\in \pproducts(\ancd)$ satisfies the conditions Th-(i), (ii), and (iii). 

({\em inductive step}): We show that for any CFD \ancd\ with $\depth{\ancd} = k$, any $m\in \pproducts(\ancd)$ satisfies the conditions Th-(i), (ii), and (iii). 

Let $\ancd = (F, r, \_^\uparrow, \agrp, \acrd)$  be a CFD with $\depth{\ancd} = k$ and $m\in \pproducts(\ancd)$.  
Holding Th-(i) is clear.  Let $\ancd' = \inducD{\ancd}{k}$ 
(the upper induced diagram of \ancd\ by depth $k$, see \defref{def:induce-depth}) and $E = \{f\in F: \depth{f} = k\}$. Let also $\asol$ denote the set of solitary features in \ancd, \ie, $\asol = \asol_\ancd$.   

{\bf Th-(ii)}:

(S-1):  
There exists $m'\in \pproducts(\ancd')$ such that $\forall f\in F \setminus E: m(f) = m'(f)$.  

Since $\depth{\ancd'} = k-1$, due to the hypothesis, 

$\forall f\in \asol \cap \chld{r} \setminus E, \exists c\in \acrd(f), \exists n'\in \pproducts(\inducN{\ancd'}{f}), ~ \forall e\in \dom{n'}: m'(e) = c\times n'(e)$. 

Due to S-1 and the fact that $(F, r, \_^\uparrow)$ is a tree of features,

(S-2): $\forall f\in \asol \cap \chld{r} \setminus E, \exists c\in \acrd(f), \exists n'\in \pproducts(\inducN{\ancd'}{f}), ~ \forall e\in \dom{n'}: m(e) = c\times n'(e)$.  

  Consider an arbitrary feature $f\in \asol \cap \chld{r} \setminus E$. There are unique $c\in \acrd(f)$ and $n'\in \pproducts(\inducN{\ancd'}{f})$  satisfying (S-2).\footnote{~$n'$ and $c$ in (S-2) are  unique multiset and multiplicity, respectively, for a given $f \in \asol \cap \chld{r} \setminus E$ satisfying the statement.} 

We define a multiset $n''$ as follows:   $n'' = n' \uplus \lceil e^i: (e\in E\cap\asol) \wedge (\prnt{e}\in \dom{n'}) \wedge (i =  m(e) / c) \rceil$. 
According to (S-2), $\forall e\in \dom{n''}: m(e) = c\times n''(e)$.

According to Def-(ii) and (iii)\footnote{ Def-(i), (ii), and (iii) stand  for \defref{def:plain-product}(i), (ii), and (iii), respectively.} and the assumption that $n'$ is a flat product of $\inducN{\ancd'}{f}$, there exists $n\in \inducN{\ancd}{f}$ such that  $\forall e\in (F\setminus E) \cup (E \cap \asol) : n''(e) = n(e)$.

Therefore, according to above and (S-2),  

$\forall f\in \asol \cap \chld{r}, \exists c\in \acrd(f), \exists n\in \pproducts(\inducN{\ancd}{f}), ~ \forall e\in \dom{n}: m(e) = c\times n(e)$.

%
%
%
%
%

 Thus, Th-(ii) holds.

{\bf Th-(iii)}: 

Let $G\in \agrp \cap 2^{\chld{r}}$. We show that $\exists n \in \pproducts(\ancd, G),~ \forall e\in \dom{n}: m(e) = n(e)$. There are the two following cases: 

(a) $k  = \depth{\ancd} > 2$,

(b) $k = \depth{\ancd} = 2$.

In the former case, $G \in \agrp_{\ancd'} \cap 2^{\chld{r}}$.  According to S-1, there exists $m' \in \pproducts(\ancd')$ such that $\forall f\in F \setminus E: m(f) = m'(f)$.

Since $\depth{\ancd'} = k-1$, due to the hypothesis, 

$\exists n' \in \pproducts(\ancd', G), \forall e\in \dom{n'}: n'(e) = m'(e)$. 

Let $n = \big(\biguplus_{f\in X} \lceil f^{m(f)}\rceil \big) \uplus n'$, where $X = E \cap \dom{m} \cap \{\gchld{f}: f\in G\}$.  

Clearly, $n\in \pproducts(\ancd, G)$. 

Since $\dom{n'} \cap E = \varnothing$ and $\forall f\in F\setminus E: m(f) = m'(f)$, we get to 

$\forall e\in \dom{n}: n(e) = m(e)$.  Thus, Th-(iii) holds in case (a).\\

Now, consider the case (b), where $\depth{\ancd} = 2$. In this case, $G\subseteq E$. 

Let $\dom{m} \cap G = \{f_1, \ldots, f_j\}$ for some $j$. 

Let $n = \biguplus_{1 \leq i \leq j} \lceil f_i ^{m(f_i)} \rceil$. 

Due to Def-(iii), $\forall 1 \leq i \leq j: m(f) \in \acrd(f)$:  (1)

Since  $f_i$ is a leaf node in $\ancd$ for any $1 \leq i \leq j$, $\lceil f_i \rceil \in \pproducts(\inducN{\ancd}{f_i})$: (2)

Due to Def-(iv), $j = |\dom{m} \cap G| \in \acrd(G)$: (3)

(1), (2), and (3) together imply that  $n \in \pproducts(\ancd, G)$. Since $\forall e\in \dom{n}: m(e) = n(e)$, Th-(iii) holds in case (b) too.

%
%
%
%
%
%
%
%
%
%
%
%

{\em \underline{Proof of (2)}}: \\
Assume that a multiset $m$ over the set of features satisfies Th-(i), (ii), (iii). We show that it also satisfies Def (ii), (iii), and (iv).   

{\bf Def-(ii)}: Recall that Def-(ii) says that $\forall f\in F_{-r}: f \in \dom{m} \implies (\exists c \in \acrd(f): m(f) = c \times m(\prnt{f}))$. 

Let $f\in F_{-r}$ and $f\in \dom{m}$. Then, either $f\in \asol$ or $\exists G\in \agrp: f\in G$. 

Let us first consider the case $f\in \asol$:  Th-(ii) implies that  there exists $c\in \acrd(f)$ and $n\in \pproducts(\inducN{\ancd}{f})$  such that $\forall e\in \dom{n}: m(e) = c\times m(\prnt{f})\times n(e)$. 
 Since $T = (F, r, \_^\uparrow)$ is a  tree of features, $m(f) = n(f) \times c \times m(\prnt{f})$. Note that $f$ is the root feature of $\inducN{\ancd}{f}$, which means that, according to \defref{def:plain-product}, $n(f) = 1$. Thus, $m(f) = c \times m(\prnt{f})$ and Def-(ii) holds.

Now, let us consider the latter case, \ie, $\exists G\in \agrp: f\in G$. Consider such a $G$ and let $G = \{f_1, f_2, \ldots, f_k\}$ for some $k$ such that $f_1 = f$. 

Th-(iii) and Th-(ii) together imply that there exists $n\in \pproducts(\ancd, G)$ such that $n^{m(\prnt{f})} \subseteq m$. 

According to \defref{def:gp-products}, there exist $c\in \acrd(G)$, $c_i \in \acrd(f_i)$, $g_i \in \{0,1\}$, and $m_i\in \pproducts(\inducN{\ancd}{f_i})$ such that 
$n = \biguplus_{1\leq i \leq k}{m_i^{c_i\times g_i}}, \text{ and } \sum_{i}{g_i} = c$. 

Since $f\in \dom{m}$, $g_1$ must be 1. (Note that $\ancd$ is an unlabelled tree of features.) 
Thus, $n(f) = m_1(f) \times c_1$.  

Since $f$ is the root feature of $\inducN{\ancd}{f_1}$ and $m_1\in \pproducts(\inducN{\ancd}{f_1})$, $m_1(f) = 1$. Therefore, $n(f) = c_1$. 

Since $T$ is an unlabelled tree, $m(f) = n(f) \times m(\prnt{f})$.  
Therefore, $m(f) =  c_1 \times m(\prnt{f})$ and  Def-(ii) holds. 


{\bf Def-(iii)}: Recall that Def-(iii) says that $\forall f\in \asol: 0\not\in \acrd(f)  \wedge m(\prnt{f}) > 0 \implies m(f) > 0$. 

Let $f$ be a solitary mandatory feature (\ie, $0 \not\in \acrd(f) $) and its parent is in $m$ (\ie, $m(\prnt{f}) > 0$). We want to show that $f$ is in $m$ too. 

The conditions Th-(ii) and (iii) imply that there exists $c\in \acrd(f)$ and $n\in \pproducts(\inducN{\ancd}{f})$ such that  $\forall e\in \dom{n}: m(e) = c\times m(\prnt{f})\times n(e)$.  
Therefore, $m(f) = n(f) \times c \times m(\prnt{f})$, as $f\in \dom{n}$.  
 
 Since $f$ is the root feature of $\inducN{\ancd}{f}$, $n(f) =1$ and $m(f) = c \times m(\prnt{f})$. 
 
 Since $ 0 \not\in \acrd(f)$ and so $m(\prnt{f}) > 0$, $m(f) > 0$. Def-(iii) holds.

{\bf Def-(iv)}: Recall that Def-(iv) says that $\forall G\in \agrp: (m(\prnt{G}) > 0) \implies (|\dom{m} \cap G| \in \acrd(G))$.

 Consider an arbitrary group $G = \{f_1, f_2, \ldots, f_k\}$ with $m(\prnt{G}) > 0$. 
 
The conditions Th-(ii) and (iii) imply that  there exists $n\in \pproducts(\ancd, G)$ such that $\forall e\in \dom{n}: m(e) = m(\prnt{G})\times n(e)$.   
 
 According to \defref{def:gp-products}, there exist $c\in \acrd(G)$, $c_i \in \acrd(f_i)$, $g_i \in \{0,1\}$, and $m_i\in \pproducts(\inducN{\ancd}{f_i})$ such that  $n = \biguplus_{1\leq i \leq k}{m_i^{c_i\times g_i}}, \text{ and } \sum_{i}{g_i} = c$. 
 
 The condition  $\sum_{i}{g_i} = c$ implies that  $|\dom{m} \cap G| \in \acrd(G)$. Hence, Def-(iv) holds. 
\end{proof}
~\\
~\\
{\bf Theorem \ref{th:unique}.}
Given two CFDs \ancd\ and $\ancd'$, $\products(\ancd) = \products(\ancd') \implies \ancd = \ancd'$. 
\begin{proof}
Let $\ancd = (F, r, \_^\uparrow, \agrp, \acrd)$ and $\ancd' = (F', r', \_^{\uparrow'}, \agrp', \acrd')$  be two CFDs such that $\products(\ancd) = \products(\ancd')$. \\

Obviously, $\bigcup_{m\in \products(\ancd)} \dom{\pure{F}{m}} = F$ and $\bigcup_{m'\in \products(\ancd')}$ $\dom{\pure{F'}{m'}} = F'$. Since $\products(\ancd) = \products(\ancd')$, $F= F'$. (S-1)\\

We give an inductive reasoning based on $\depth{\ancd}$ (the depth of \ancd) to show that $\ancd = \ancd'$.  \\


({\em base case}): Let $\ancd = 1$, \ie, $F = \{r\}$,  and $\prnt{\_} = \agrp = \acrd = \varnothing$. According to (S-1), $F' = \{r\}$, which implies that $r' = r$, $\_^{\uparrow'} =  \agrp' = \acrd' = \varnothing$. Therefore, $\ancd = \ancd'$. \\

({\em hypothesis}): Assume that for some $n\in \nat$ and for any $\depth{\ancd} < n$: $\products(\ancd) = \products(\ancd') \implies \ancd = \ancd'$. \\

({\em inductive step}):  We want to show that if $\depth{\ancd} =n$, then $\ancd = \ancd'$. 

Let us suppose that $r$ in \ancd\ ($r'$ in $\ancd'$, respectively) has $k$ ($x$, respectively) solitary subfeatures $f_1, \ldots, f_k$ ($f'_1, \ldots, f'_x$, respectively)  and $t$ ($y$, respectively) groups $\{G_1, \ldots, G_t\}$ ($\{G'_1, \ldots, G'_y\}$, respecively). According to \defref{def:hier-products},   

$$\products(\ancd) = \{\lceil r, m_1^{c_1}, \ldots, m_k^{c_k}, g_1, \ldots, g_t\rceil, \text{ where}$$ 
$\forall 1\leq i \leq k, \forall 1\leq j \leq t:$  

$m_i\in \products(\inducN{\ancd}{f_i}), c_i \in \acrd(f_i), g_j \in \products(\ancd, G_j)\}$ (C)

$$\products(\ancd') = \{\lceil r', m_1^{c_1}, \ldots, m_{x}^{c_x}, g_1, \ldots, g_y \rceil, \text{ where}$$ 
$\forall 1\leq i \leq x, \forall 1\leq j \leq y:$ 

$m_i\in \products(\inducN{\ancd'}{f'_i}), c_i \in \acrd'(f'_i), g_j \in \products(\ancd', G'_j)\}$. (C') \\

Consider an arbitrary hierarchical product $m = \lceil r, m_1^{c_1}, \ldots, m_k^{c_k}, g_1, \ldots, g_t\rceil$, where $m_i$ ($1\leq i \leq k$) and $g_j$ ($1\leq j \leq t$) satisfy the conditions in (C). Since for any $1\leq i \leq k$ and $1\leq j \leq t: \rankm{m_i} \in \multhier(F) \wedge  \rankm{g_j} \in \multhier(F)$, $r$ is the only urelement in the domain of $m$, \ie, $m\in \products(\ancd'): \dom{m'} \cap F' = \{r'\}$.  Likewise, for any $m\in \products(\ancd'): \dom{m'} \cap F' = \{r'\}$. Since $\products(\ancd) = \products(\ancd')$, $r = r'$.


For any CFD, the domain of any hierarchical product of an induced diagram by a node $f$ includes  $f$ with multiplicity $1$ and its all other elements are multisets. Also, the domain of a grouped hierarchical product of a CFD is a set of multisets, \ie, it  does not include any urelement. This implies the following statements:

(i)  $k = x$ and $t = y$,

(ii)  $\forall 1\leq i \leq k, \exists 1\leq j \leq k:   \products(\inducN{\ancd}{f_i}) = \products(\inducN{\ancd'}{f'_{j}}) \wedge \acrd(f_i) = \acrd'(f'_j)$,

(iii) $\forall 1\leq i \leq t, \exists 1\leq i' \leq t:   \products(\ancd, G_i) = \products(\ancd', G'_j)$.
~\\

(ii) implies that the sets of $r$'s solitary subfeatures in both \ancd\ and $\ancd'$ are the same. Without loss of generality, suppose that $\forall 1\leq i\leq k: f_i = f'_i$. Since  $\forall 1\leq i\leq k: \products(\inducN{\ancd}{f_i}) = \products(\inducN{\ancd'}{f'_i})$ and $\depth{\inducN{\ancd}{f_i}} < n$, due to the hypothesis, $\inducN{\ancd}{f_i} = \inducN{\ancd'}{f_i}$. \\

(iii) implies that the set of groups of $r$ in \ancd\ and $\ancd'$ are the same. Without loss of generality, we suppose that $\forall 1\leq i \leq t: G_i = G'_i$. Consider an $1\leq i \leq t$ and let $G_i = G'_i = \{q_1, \ldots, q_z\}$. According to \defref{def:g-products}, \\

$\products(\ancd, G_i) = \{ \lceil m_1^{c_1\times l_1}, \ldots, m_z^{c_z \times l_z}\rceil: \forall 1\leq j \leq z.~ m_j \in \products(\inducN{\ancd}{g_j}), c_j\in \acrd(g_j), l_j \in \{0,1\}, \text{ and }  l_1 + \ldots + l_z \in \acrd(G_i) \}$.\\

$\products(\ancd', G_i) = \{ \lceil m_1^{c_1\times l_1}, \ldots, m_z^{c_z \times l_z}\rceil: \forall 1\leq j \leq z.~ m_j \in \products(\inducN{\ancd'}{g_j}), c_j\in \acrd'(g_j), l_i \in \{0,1\}, \text{ and } l_1 + \ldots + l_z \in \acrd'(G_i) \}$. \\

$\products(\ancd, G_i) = \products(\ancd', G_i)$ implies that $\forall 1\leq j \leq z: \products(\inducN{\ancd}{g_j}) = \products(\inducN{\ancd'}{g_j})$ and $\acrd(g_j) = \acrd'(g_j), \acrd(G_i) = \acrd'(G_i)$. Since $\depth{\inducN{\ancd}{g_i}} < n$, due to the hypothesis, $\inducN{\ancd}{g_i} = \inducN{\ancd'}{g_i}$. \\

According to above, since $r$ in both \ancd\ and $\ancd'$ have the same set of solitary subfeatures and groups whose corresponding induced diagrams are the same with the same multiplicities,  $\ancd = \ancd'$.  
\end{proof}
~\\
~\\
{\bf \theoref{th:hier-flat}.}
For any CFD $\ancd\in \cfdfam(F)$, the function $\restr{\xxx{flat}_F}{\products(\ancd)}$, \ie, the restriction of $\xxx{flat}_F$ to the subdomain $\products(\ancd)$, provides a bijection between $\products(\ancd)$ and $\pproducts(\ancd)$.
\begin{proof}
 We use an inductive reasoning based on the depth of CFDs to show this. 

({\em base case}): The statement obviously holds  for any CFD with singleton tree, \ie, a CFD with depth 1. 

({\em hypothesis}): Assume that the statement holds for any CFD $\ancd$ with $1 \leq \depth{\ancd} < d$ for some $d\in \nat$. 

({\em inductive step}): We show that $\restr{\xxx{flat}_F}{\products(\ancd)}$ provides a bijection from $\products(\ancd)$ to $\pproducts(\ancd)$ for any CFD \ancd\ with $\depth{\ancd} = d$. 

Let $\ancd = (F, r, \prnt{\_}, \agrp, \acrd)$ be a CFD with $\depth{\ancd} = d$ and $\asol \subseteq F_{-r}$ denote the set of its solitary features. Suppose that $\asol \cap \chld{r} = \{f_1, \ldots, f_i\}$ (solitary subfeatures of the root) and $\agrp \cap 2^{\chld{r}} = \{G_1, \ldots, G_j\}$ (groups subelements of the root) for some $i,j \in \nat$. 

Consider a hierarchical product $h\in \products(\ancd)$.  According to \defref{def:hier-products}, $h$ is a multiset $\lceil r, h_1^{c_1}, \ldots, h_i^{c_i}, g_1, \ldots, g_j \rceil$, where $h_k \in \products(\inducN{\ancd}{f_k})$, $c_k \in \acrd(f_k)$ ($1\leq k \leq i$), and $g_t \in \products(\ancd, G_t)$ ($1\leq t \leq j$). 

According to \defref{def:flatten}, $\pure{F}{h} = \lceil r \rceil \uplus \biguplus_{1\leq k \leq i} (\pure{F}{h_k})^{c_k} \uplus \biguplus_{1\leq t\leq j} \pure{F}{g_t}.$

Since $h_k \in \products(\inducN{\ancd}{f_k})$ and $\depth{\inducN{\ancd}{f_k}} < d$ for $1\leq k \leq i$, due to the hypothesis, $\pure{F}{h_k}$ is a  flat product of the diagram induced by $f_k$, \ie, $\pure{F}{h_k}\in \pproducts(\inducN{\ancd}{f_k})$. 

Let $G_t = \{g_1, \ldots, g_l\}$ for $1\leq t \leq j$. 

According to \defref{def:g-products}, ${g_t = \lceil m_1^{c'_1\times t_1}, \ldots, m_l^{c'_n\times t_l}\rceil}$, where $m_k\in \products(\inducN{\ancd}{g_k})$, $c'_k \in \acrd(g_k)$, $t_k \in \{0,1\}$, and $t_1 + \ldots +  t_l \in \acrd(G_t)$ ($1\leq k \leq l$). According to \defref{def:flatten}, $\pure{F}{g_t} = \pure{F}{m_1}^{c'_1\times t_1} \uplus \ldots \uplus \pure{F}{m_l}^{c'_n\times t_n}$. Since $\depth{\inducN{\ancd}{g_k}} < d$ for any $1\leq k \leq l$, due the hypothesis, $\pure{F}{m_k} \in \pproducts(\inducN{\ancd}{g_k})$.  This implies that, according to \defref{def:gp-products}, $\pure{F}{g_t} \in \pproducts(\ancd, G_t)$. 

According to above, $\pure{F}{h} = \lceil r \rceil \uplus \biguplus_{1\leq k \leq i} m_k^{c_k} \uplus \biguplus_{1\leq t\leq j} n_t$, where $m_k = \pure{F}{h_k}$ ($1\leq k \leq i$) is a  flat product of the diagram induced by $f_k$, \ie, $m_k\in \pproducts(\inducN{\ancd}{f_k})$  and $n_t$ ($1\leq t\leq j$) is a flat grouped product of $G_t$, \ie, $n_t = \pure{F}{g_t} \in \pproducts(\ancd, G_t)$. Due to \lemref{th:flat-products}, $\pure{F}{h} \in \pproducts(\ancd)$. Therefore, $\restr{\xxx{flat}_F}{\products(\ancd)}$ maps each hierarchical product of \ancd\  to a flat product of \ancd.  In the following, we show that $\restr{\xxx{flat}_F}{\products(\ancd)}$ is an injective function.  

Consider two different hierarchical products $h, h'\in \products(\ancd)$ such that $\pure{F}{h} = \pure{F}{h'}$. According to \defref{def:flatten} and \defref{def:hier-products}, 

$\pure{F}{h} = \lceil r \rceil \uplus \biguplus_{1\leq k \leq i} \pure{F}{h_k}^{c_k} \uplus \biguplus_{1\leq t\leq j} \pure{F}{g_t}$, and 

$\pure{F}{h'} = \lceil r \rceil \uplus \biguplus_{1\leq k \leq i} \pure{F}{h'_k}^{c'_k} \uplus \biguplus_{1\leq t\leq j} \pure{F}{g'_t}$,  where  

$\forall 1\leq k \leq i, \forall 1\leq t \leq j$: $h_k, h'_k \in \products(\inducN{\ancd}{f_k})$, $c_k, c'_k \in \acrd(f_k)$, and $g'_t, g_t \in \products(\ancd, G_t)$.

Note that for any two distinct subelements (solitary and/or group subelements) of the root, their hierarchical and flat products are built on disjoint subsets of features (a CFD is a special tree of features). Therefore,  $\pure{F}{h} = \pure{F}{h'}$ implies that for any $1\leq k \leq i, 1\leq t \leq j$: $\pure{F}{h_k} = \pure{F}{h'_k}$, $c_k = c'_k$, and $\pure{F}{g_t} = \pure{F}{g'_t}$.  Due to hypothesis, this implies that for any $1\leq k \leq i, 1\leq t \leq j$: $h_k = h'_k$, $c_k = c'_k$, and $g_t = g'_t$. Therefore, $h=h'$, which implies that the restriction of the function $\restr{\xxx{flat}_F}{\products(\ancd)}$ is an injective function from $\products(\ancd)$ to $\pproducts(\ancd)$. 

According to \defref{def:hier-products} and \lemref{th:flat-products}, $|\products(\ancd)| = |\pproducts(\ancd)|$ (recursive definitions of hierarchical and flat products of \ancd) for any CFD $\ancd$, \ie, the cardinalities of the sets of flat and hierarchical products of \ancd\ are the same. Therefore, the restriction of the flattening function to the hierarchical semantics of \ancd\ is a surjective function, as it is injective and the cardinalities of the domain and codomain are the same. 

According to above, $\restr{\xxx{flat}_F}{\products(\ancd)}: \products(\ancd) \rarr \pproducts(\ancd)$ is a bijection.  
\end{proof}

~\\
{\bf Theorem \ref{th:hierp-2-treemult}.}
Any hierarchical product of a given CFD over a set of features $F$ is a tree-like multiset over $F$. 
\begin{proof} 
We use an inductive reasoning based on the depth of CFDs to deal with this theorem. 

({\em base case}): Obviously, the statement holds for any CFD \ancd\ with $\depth{\ancd} = 1$. 

({\em hypothesis}): We assume that for any CFD \ancd\ with  $1 \leq \depth{\ancd} < n$,  the statement holds. 

({\em inductive step}): Let $\ancd = (F, r, \_^\uparrow, \agrp, \acrd)$ be a CFD  and $\depth{\ancd} = n$. We show that any hierarchical product $m\in \products(\ancd)$ is a tree-like multiset.

Consider the CFD $\ancd' \eqdef \inducD{\ancd}{n}$ (upper diagram Induced by depth $n$). Due to \defref{def:hier-products}, for any hierarchical product $m\in \products(\ancd)$, there exists $m'\in \products(\ancd')$ such that $m$ is obtained by replacing any feature $f \in \{f\in F: \depth{f} = n-1\}$ in $m'$ with an $x\in \products(\inducN{\ancd}{f})$. 

Due to the hypothesis, any $x\in \products(\inducN{\ancd}{f})$ is a tree-like multiset. Thus, according to \defref{def:tree-mult}, $m$ would be a tree-like multiset. 
\end{proof}
~\\
~\\
{\bf \theoref{th:treemult-2-CFD}.}
For any tree-like multiset $t$, there is a CFD $\ancd$ such that $t\in \products(\ancd)$. 
\begin{proof}
\label{pp:proof-treemult-2-CFD}
Let $t$ be a tree-like multiset. We want to show that there is a CFD whose hierarchical semantics includes $t$. 

Let $T =  (N, r, \prnt{\_})$ and $\agrp$  denote the tree and groups  associated with $t$, respectively: $N = N_t$, $r= r_t$, $\prnt{\_} = {\_}^{\uparrow_t}$, and $\agrp = \agrp_t$.\footnote{~See Definitions \ref{def:tree-treemult} and  \ref{def:groups-treemult}, respectively.}  We also define a function $\acrd: (N\setminus \{r\} \cup \agrp) \rarr 2^\nat$ as follows. $\forall e\in (N\setminus \{r\} \cup \agrp): \acrd(e) = \{\acrd_t(e)\}$, where $\acrd_t: (N\setminus \{r\}) \cup \agrp$ is defined in \defref{def:card-treemult}. 

The tuple $\ancd = (T, \agrp, \acrd)$ would be a CFD except that there may be some singleton groups (note that singleton groups are not allowed in CFDs--see \defref{def:CFD-mult}(ii)). Let us call a CFD in which singleton groups are allowed a {\em CFD plus} (\cfdplus). The semantics of \cfdplus s can be defined via hierarchical semantics of CFDs. Note that the definition of hierarchical semantics for CFDs (\defref{def:hier-products}) can be directly used on \cfdplus s. In this sense, the tuple $(T, \agrp, \acrd)$ represents a singleton hierarchical semantics, as all multiplicities are singleton. It is easy to see that its singleton hierarchical product is $t$. 

Thus, \ancd\ is a CFD plus representing $t$ as its single hierarchical product. We show that this tuple is a substructure of some CFDs. Indeed, to get a CFD whose hierarchical semantics includes the single hierarchical product of the tuple, we just need to add one (or more than one) feature(s) to singleton groups. We formally show how this works in the following. 

Let $\agrp^1 = \{ G\in \agrp: |G| = 1\}$ and $N'$ be a set of symbols (features) with $N' \cap N = \varnothing$ and $|N'| = |\agrp^1|$. Consider a  bijection $l: \agrp^1 \rarr N'$.  We build a CFD $\ancd' = (N' \cup N, r, {\_}^{\uparrow'}, \agrp', \acrd')$ as follows.

$$
\forall n\in N \cup N': 
\uparrow'(n) = 
\begin{cases}
\prnt{l^{-1}(n)} & \text{if } n\in N' \\
\prnt{n} & \text{otherwise} 
\end{cases}  
$$
$$
\agrp' = (\agrp\setminus \agrp^1) \cup \{G \cup \{l(G)\}: G\in \agrp^1\}
$$  
$\acrd': ((N' \setminus \{r\}) \cup \agrp') \rarr 2^\nat$ is defined as follows. 
$$
\forall e\in (N' \setminus \{r\}) \cup \agrp'):
\acrd'(e) = 
\begin{cases}
\acrd(e) & \text{if } e\in N \vee e\in \agrp \setminus \agrp^1\\
\{1\} & \text{otherwise} 
\end{cases}  
$$
Clearly, $\ancd'$ is a CFD and $\ancd$ is a substructure of $\ancd'$. Thus, $t\in \products(\ancd')$. The theorem is proven! 
\end{proof}
~\\
~\\
{\bf Theorem \ref{th:merg-2treemults}.} 
Consider an enumerable set of tree-like multisets $U = \{t_i: i\in I\} \subset \treemulthier(A)$ over a set $A$, where $I$ enumerates its elements. 
  Let $T_i = (N_i, r_i, {\_}^{\uparrow_i})$ and $\agrp_i$  ($\forall i\in I$) denote the $t_i$'s associated tree and groups, respectively (see Definitions \ref{def:tree-treemult} and \ref{def:groups-treemult}, respectively).   The tree-like multisets in $U$ are mergeable iff:

(i) $\forall i, j \in I: T_i, T_j$ are mergeable.

(ii) $\forall i,j\in I, \forall n\in N_i\cap N_j: (\exists G\in \agrp_i: n\in G) \implies (\exists G\in \agrp_j: n\in G)$.
\begin{proof} \label{proof:merg-2treemults}
We prove the statement for $I = \{1,2\}$. The proof can be easily extended to any enumerating set $I\subseteq \nat$. 
Let $U = \{t_1, t_2\}$. We need to show that the following statements hold:

~~~~(1) $t_1$ and $t_2$ are mergeable $\implies$ (i) and (ii) hold.

~~~~(2) (i) and (ii) hold $\implies$ $t_1$ and $t_2$ are mergeable. 

{\em \underline{Proof of (1)}}: 

Suppose that $t_1$ and $t_2$ are megeable.  According to \defref{def:mergeable-treemult}, there exists a CFD $\ancd = (T, \agrp, \acrd)$ with $T = (F, r, \prnt{\_})$ such that  $t_1, t_2 \in \products(\ancd)$.  This implies the following statements:

(S-1) $T_1$ and $T_2$ are subtrees of $T$ such that  their roots are equal to the root of $T$.  Formally, $N_1 \cup N_2 \subseteq F$, $r_1 = r_2 = r$, and $\forall n \in N_1 \cap N_2 \setminus \{r\} : {n}^{\uparrow_1} = {n}^{\uparrow_2} = \prnt{n}$. Thus (i) holds.

(S-2) For any urelement $a\in A$, if its corresponding induced tree in $t_1$ (\ie, $\inducN{t_1}{a}$) or $t_2$ (\ie, $\inducN{t_2}{a}$) is a grouped tree-like multiset, then $a$ must be a grouped feature in \ancd. Formally, $\forall a\in A: (\exists G\in \agrp_1: a \in G) \vee (\exists G\in \agrp_2: a \in G) \implies (\exists G\in \agrp: a \in G)$. Clearly, this implies that $\forall n\in N_1\cap N_2: (\exists G_1\in \agrp_1: n\in G_1) \implies (\exists G_2\in \agrp_2: n\in G_2)$. Therefore,  (ii) holds. 


Due to (S-1) and (S-2), (1) is proven. 

{\em \underline{Proof of (2)}}:

Suppose that (i) and (ii) hold. We show that $t_1$ and $t_2$ are mergeable. To this end, we construct a CFD   whose hierarchical semantics includes both $t_1$ and $t_2$.  


Let $N' = N_1 \cup N_2$, $r'= r_1$ (note that $r_1 = r_2$), and $\_^{\uparrow'}: N'\setminus \{r'\} \rarr N'$ defined as  $\_^{\uparrow'} =  \_^{\uparrow_1} \cup \_^{\uparrow_2}$. Note that $(N', r', \_^{\uparrow'}) = \merg{\{T_1, T_2\}}$ (see \defref{def:merge-tree-treemult}). 

 
 Let  $\agrp' = \agrp_\sqcap  \bigcup \agrp_\sqcup  \text{, where }$

 ~~~~~~~~$\agrp_\sqcap = \{G_1 \cup G_2: (G_1\in \agrp_1) \wedge (G_2 \in \agrp_2) \wedge (G_1 \cap G_2 \neq \varnothing)\}$,

 ~~~~~~~~$\agrp_\sqcup = \{G \in \agrp_1 \cup \agrp_2: (\forall G'\in \agrp_\sqcap: G' \cap G = \varnothing)\}$,

To merge two CFDs, we also need to merge their groups. According to the definition of CFDs, two different groups in a CFD must share no elements. Thus, we have to merge all groups in $\agrp_1$ and $\agrp_2$ that share some elements. $\agrp_\sqcap$ does so. Any other groups in either $\agrp_1$ and $\agrp_2$ must have to be considered as a group in the merged CFD. Such groups are obtained via $\agrp_\sqcup$. There may be some singleton elements in $\agrp$. Note that, according to \defref{def:CFD}, singleton groups are not allowed in a CFD. Below, we address this problem. 

Let ${\agrp''} = \{G\in \agrp: |G| = 1\}$  and $N''$ be a set of symbols (features) with $N'' \cap N' = \varnothing$ and $|N''| = |\agrp''|$. Consider a  bijection $l: \agrp'' \rarr N''$. 

We define a  tuple $\ancd = (N, r, \prnt{\_}, \agrp, \acrd)$,  where: 

$N = N'' \cup N',$

$r = r',$

$\agrp = (\agrp' \setminus \agrp'') \cup \{G \cup \{l(G)\}: G\in \agrp''\},$  

$\prnt{\_} : N\setminus \{r\} \rarr N$, defined as: 
$$
\forall n\in N: 
\prnt{n} = 
\begin{cases}
{l^{-1}(n)}^{\uparrow'} & \text{if } n\in N'' \\
n^{\uparrow'} & \text{otherwise} 
\end{cases}  
$$
\begin{equation*}
\forall e\in N \cup \agrp: \acrd(e) = 
\begin{cases}
\{0\} \cup \acrd_{1}(e) & \text{if } (e\in N_{1} \setminus N_{2}) \vee (e\in \agrp_\sqcup \cap \agrp_{1})\\
\{0\} \cup \acrd_{2}(e) & \text{if } (e\in N_{2} \setminus N_{1}) \vee (e\in \agrp_\sqcup \cap \agrp_{2})\\
\acrd_{1}(e)  \cup \acrd_{2}(e) & \text{if } (e\in N_{1} \cap N_{2}) \vee (e\in \agrp_\sqcap)\\
\{1\} & \text{otherwise}
\end{cases}  
\end{equation*}
where  $\acrd_1$ and  $\acrd_2$  denote  the multiplicities associated with  $t_1$  and  $t_2$, respectively (see \defref{def:card-treemult}).

%
%
%
%
%

%
%
%
%
%
%
It is easy to see that the tuple $\ancd = (N, r, \prnt{\_}, \agrp, \acrd)$ is a CFD. It is obvious that $t_1$ and $t_2$ are two hierarchical products of $\ancd$. Thus, $t_1$ and $t_2$ are two mergeable tree-like multisets. \\

The proof is easily extendable to any enumerating set $I\subseteq \nat$, as a set of tree-like multisets are mergeable iff each pairs of tree-like multisets are mergeable. 
\end{proof}
~\\
~\\
{\bf Theorem  \ref{th:merge-relax-yn}.}
Consider an emeumrable set of tree-like multisets $U \subset \treemulthier(A)$ over a set $A$ of urelements.  

(i) $U$ is mergeable iff $\relaxm{U}$ is. 

(ii) $U$ is mergeable implies that $\relaxm{U}$ is finite. 
\begin{proof} [{\bf Proof of  \theoref{th:merge-relax-yn}}] 
Let $U = \{t_i: i\in I\} \subset \treemulthier(A)$, where $I \subseteq \nat$ enumerates the elements of $U$. Let $T_i = (N_i, r_i, {\_}^{\uparrow_i})$, $\agrp_i$, and $\acrd_i$, for any $i\in I$,  represent the $t_i$'s associated tree, groups, and multiplicities , respectively -- see Definitions \ref{def:tree-treemult}, \ref{def:groups-treemult}, and \ref{def:card-treemult}. 

{\em \underline{Proof of (i)}}: 

For any $i\in I$, let $\relaxm{T_i}$ and $\relaxm{\agrp_i}$ 
 denote the tree and groups associated with $\relaxm{t_i}$ (the relaxed multiset of $t_i$). According to \propref{prop:relax-cfd}, $ \forall i\in I: \relaxm{\agrp_i} = \agrp_i \text{ and } \relaxm{T_i} = T_i$. 
 
 
According to \theoref{th:merg-2treemults},  

$U$ is mergeable 

~~~~~~~~~~~~ $\biimplies$ 

~~~~~ -- $\forall i, j \in I: T_i, T_j$ are mergeable. 

~~~~~ -- $\forall i,j\in I, \forall n\in N_i\cap N_j: (\exists G\in \agrp_i: n\in G) \implies (\exists G\in \agrp_j: n\in G)$.

~~~~~~~~~~~~ $\biimplies$

~~~~~ -- $\forall i, j \in I: \relaxm{T_i}, \relaxm{T_j}$ are mergeable. 

~~~~~ -- $\forall i,j\in I, \forall n\in N_i\cap N_j: (\exists G\in \relaxm{\agrp_i}: n\in G) \implies (\exists G\in \relaxm{\agrp_j}: n\in G)$.

~~~~~~~~~~~~ $\biimplies$

According to \theoref{th:merg-2treemults},  $\relaxm{U}$ is meageable. 

%
%
%
%
%
%
%
%

{\em \underline{Proof of (ii)}}: 

Suppose that the elements of $U$ are megeable. Let $\ancd \in \mergcfd{U}$, \ie, \ancd\ is a minimal representative CFD of $U$. Let $\ancd = (T, \agrp, \acrd)$ with $T = (N, r, \prnt{\_})$.   

According to (i), the elements of $\relaxm{U}$ are megeable. Recall that the only difference between a tree-like multiset and its relaxed multiset is in their multiplicities, \ie, their trees and groups would be the same. We build a representative CFD $\relaxm{\ancd}$ of $\relaxm{U}$, as follows:  

$\relaxm{\ancd} = (T, \agrp, \relaxm{\acrd})$ where 
$$
\forall e\in (N \setminus \{r\}) \cup \agrp:
\relaxm{\acrd}(e) = 
\begin{cases}
\acrd(e) & \text{if } e\in \agrp \\
\{0,1\} & \text{otherwise} 
\end{cases}   
$$
Clearly, $\relaxm{\ancd}$  is a  representative CFD of $\relaxm{U}$, since $\ancd$ is a minimal representative CFD of $U$ and all feature multiplicities in $\relaxm{\ancd}$ are $\{0,1\}$.  Since there is no feature in $\relaxm{\ancd}$ with an infinite  multiplicity domain, $\products(\relaxm{\ancd})$ would be finite. Thus, $\relaxm{U}$ is finite, since $U\subseteq \products(\relaxm{\ancd})$. 
%
%
\end{proof}
~\\
~\\
{\bf \theoref{th:charc-complete-merge}.}
Consider an emeumerable set of tree-like multisets $U \subset \treemulthier(A)$ over a set $A$ of urelements. $U$ is completely mergeable iff

(i) $\relaxm{U}$ is completely mergeable, and

(ii) $\forall t\in \relaxm{U}, \forall a\in \dom{\pure{A}{t}}, \forall c\in \acrd_{U}(a), \exists t'\in U: (\relaxm{t'} = t) \wedge (\#_{t'}(\inducN{t'}{a}) = c)$. 
\begin{proof}
Suppose that $U$ is completely mergeable, which means that there is some CFD $\ancd = (T, \agrp, \acrd)$ with $(F, r, \prnt{\_})$ representing $U$. We want to show that the statements (i) and (ii) hold. 

We build a CFD $\relaxm{\ancd} = (T, \agrp, \relaxm{\acrd})$, where $\relaxm{\acrd}: (F \setminus \{r\}) \cup \agrp \rarr 2^\nat$ is defined as follows:
$$
\forall e\in (F \setminus \{r\}) \cup \agrp:
\relaxm{\acrd}(e) = 
\begin{cases}
\acrd(e) & \text{if } e\in \agrp \\
\{0,1\} & \text{if } (e \not\in \agrp) \wedge (0 \not\in \acrd(e)) \\
\{1\} & \text{if } (e \not\in \agrp) \wedge (0 \not\in \acrd(e))
\end{cases}   
$$
It follows obviously that $\products(\relaxm{\ancd}) = \relaxm{U}$. Therefore, $\relaxm{U}$ is completely mergeable. 

Now, consider a tree-like multiset $t\in \relaxm{U}$, $a\in \dom{\pure{A}{t}}$, and $c\in \acrd_{U}(a)$. We want to show that there exists $t'\in U$ such that $\relaxm{t'} = t$ and $\#_{t'}(\inducN{t}{a}) = c$.  

$t\in \relaxm{U}$ implies that $t\in \products(\relaxm{\ancd})$. $a$ is a feature in $\relaxm{\ancd}$ involved in $t$ and $c$ is a valid multiplicity of the feature $a$ in $\ancd$ (see the definition of overall multiplicities in \defref{def:overallmult}).  

Since $t\in \relaxm{U}$, there is some $t''\in U$ such that $\relaxm{t''} = t$. If $\#_{t''}(\inducN{t''}{a}) = c$, then the statement (ii) is proven. Suppose that $\#_{t''}(\inducN{t''}{a}) \neq c$. $t''$ is a hierarchical product of $\ancd$. Thus, for any $f\in \dom{\pure{A}{t}}$ (including $a$), $\#_{t''}(\inducN{t''}{f}) \in \acrd(f)$.  According to \defref{def:hier-products}, replacing $\#_{t''}(\inducN{t''}{f})$ by any other valid multiplicity in the multiplicity domain of $f$ would give us another valid hierarchical product of $\ancd$. Let us define $t'$ by replacing $\#_{t''}(\inducN{t''}{a})$ by $c$. $t' \in \products(\ancd)$ and thus $t\in U$. The statement (ii) is proven. \\

Proving that $U$ is completely mergeable if the statements (i) and (ii) hold is very straightforward: Suppose that (i) and (ii) hold. Therefore, there exists a CFD $\relaxm{\ancd}$ such that $\products(\relaxm{\ancd}) = \relaxm{U}$. Note that the multiplicity domain of any feature in $\relaxm{\ancd}$ is either $\{0,1\}$ or $\{1\}$.  Now, we define a CFD $\ancd$ by replacing the multiplicity of any feature $a$ in $\relaxm{\ancd}$ by $\acrd_{U}(a)$ (overall multiplicity og $a$, see \defref{def:overallmult}). Clearly, according to (ii),  $\products(\ancd) = U$. Therefore, $U$ is completely mergeable. 
\end{proof}

\end{document}